\documentclass{aa}

\usepackage{graphicx}

\begin{document}

\newcommand{\be}{\begin{equation}}
\newcommand{\ee}{\end{equation}}

\title{An efficient shock-capturing central-type scheme \\
  for multidimensional relativistic flows}
\subtitle{II. Magnetohydrodynamics}

\author{
L. Del Zanna            \inst{1} 
\and    N. Bucciantini  \inst{1} 
\and    P. Londrillo    \inst{2}
}

\institute{
Dipartimento di Astronomia e Scienza dello Spazio,
Universit\`a degli Studi di Firenze, 
Largo E. Fermi 2, 50125 Firenze, Italy
\and
Osservatorio Astronomico di Bologna, 
Via C. Ranzani 1, 40127 Bologna, Italy
}

\offprints{L. Del Zanna; \\
\email{ldz@arcetri.astro.it}}

\date{Received ...; accepted ...}

\abstract{
A third order shock-capturing numerical scheme for three-dimensional 
special relativistic magnetohydrodynamics (3-D RMHD) is presented and
validated against several numerical tests.
The simple and efficient {\em central} scheme described in Paper~I 
(Del Zanna and Bucciantini, {\em Astron. Astrophys.}, 390, 1177--1186, 2002)
for relativistic hydrodynamics is here extended to the magnetic case 
by following the strategies prescribed for classical MHD by 
Londrillo and Del~Zanna ({\em Astrophys. J.}, 530, 508--524, 2000).
The scheme avoids completely spectral decomposition into characteristic
waves, computationally expensive and subject to many degenerate cases
in the magnetic case, while it makes use of a two-speed Riemann solver that 
just require the knowledge of the two local fast magnetosonic velocities.
Moreover, the onset of spurious magnetic monopoles, which is a typical
problem for multi-dimensional MHD upwind codes, is prevented by properly 
taking into account the solenoidal constraint and the specific antisymmetric 
nature of the induction equation. Finally, the extension to generalized 
orthogonal curvilinear coordinate systems is included, thus the scheme 
is ready to incorporate general relativistic (GRMHD) effects.

}

\authorrunning{L.~Del Zanna et al.}
\titlerunning{A shock-capturing scheme for relativistic flows -- II.}

\maketitle


\section{Introduction}

Most of the astrophysical sources of high-energy radiation and particles 
are believed to involve the presence of relativistic motions in
{\em magnetized} plasmas. For example, the radio emission from
extra galactic jets (especially from terminal radio lobes) or
from plerion-like supernova remnants is due to 
synchrotron radiation produced by relativistic electrons spiraling 
around magnetic field lines, thus indicating the presence of
significant magnetic fields.
Strong magnetic fields are supposed to play an essential role in 
converting the energy of accreting material around super-massive 
black holes at the center of Active Galactic Nuclei (AGNs), 
into powerful relativistic jets escaping along open field lines 
(Begelman et al. 1984).
Similar phenomena may be at work in the galactic compact X-ray sources
known as microquasars (Mirabel \& Rodriguez 1994).
These processes involve the interaction of relativistic gasdynamic flows 
and shocks with strong magnetic fields, which have now started
to be studied via computer simulations (see Meier et al. 2001 for a review).
Powerful relativistic blast shocks should also be at the origin
of the still mysterious gamma-ray bursts (GRBs; M\'esz\'aros \& Rees 1992).
Moreover, the presence of magnetic field has been invoked in various 
astrophysical objects to explain both their morphology and evolution by
applying simplified analytical models to basic plasma physics effects
(e.g. the magnetic pinch and kink instabilities may affect the structure
of both AGN and microquasar jets, and possibly also the overall shape
of pulsar wind nebulae), although a detailed study of the nonlinear 
and turbulent regimes is still lacking.

Due to the extreme complexity and richness of the possible effects arising 
in relativistic plasma physics, there is a very strong interest among the 
astrophysical community in the development of computer codes for both
relativistic hydrodynamics (RHD) and magnetohydrodynamics (RMHD), since
in most cases only numerical simulations are able to cope with the evolution
of such phenomena. After some early attempts based on
non-conservative schemes that handled shocks with the aid of large
artificial viscosity and resistivity, it is only during the last
decade that conservative shock-capturing Godunov-type numerical codes,
already successfully applied to gasdynamic problems, have started
to be applied to RHD too, achieving both high accuracy in smooth regions
of the simulated flow and sharp discontinuous profiles at shocks
(e.g. Marquina et al. 1992; Schneider et al. 1993; Balsara 1994; Duncan \&
Hughes 1994; Eulderink \& Mellema 1994;  Font et al. 1994; 
Dolezal \& Wong 1995; Falle \& Komissarov 1996; Donat et al. 1998;
Aloy et al. 1999; Del Zanna \& Bucciantini 2002).

However, in spite of the success of Godunov-type RHD codes and,
at the same time, of the presence of various extensions of gasdynamic
schemes to classical MHD (see the recent review by T\'oth, 2000), to date
only a couple of RMHD schemes have been described in the literature.
Both codes are second order accurate and are based on linearized
Riemann solvers (Roe matrix) in the definition of fluxes
at cell interfaces. This process involves the decomposition of
primitive variables in a set of {\em characteristic waves}, each
of them propagating a single discontinuity, and a further 
composition to obtain the numerical upwind fluxes. Moreover, a certain
amount of extra artificial viscosity is often needed to stabilize
the schemes in particular degenerate cases.
The two codes are described in Komissarov (1999a; KO
from now on), which is a truly multidimensional scheme, and in
Balsara (2001; BA from now on), the latter tested just 
against one-dimensional (1-D) {\em shock-tube} problems. 
There is actually another RMHD code, which has been extensively
used in relativistic 2-D and 3-D jet simulations 
(e.g. Koide et al. 1996; Nishikawa et al. 1998), 
later extended to general relativistic (GRMHD) effects (with given
Schwarzschild or Kerr metrics) and applied to the jet formation 
mechanism (e.g. Koide et al. 1999; 2000). However, this code cannot
be regarded as belonging to the Godunov family, since it is based on a
second order Lax-Wendroff scheme, thus with a very high level of implicit 
numerical viscosity.
Moreover, a complete set of the standard numerical tests, needed to
check the properties of any shock-capturing scheme, has never been 
published for such code, so it is difficult to comment on their results
and to compare the respective code performances (especially on
contact discontinuities, where shock-capturing codes not based on
linearized Riemann solvers are usually less accurate).

The reasons behind the difficulty of extending shock-capturing
relativistic gasdynamic codes to the magnetic case are essentially
the same encountered in building classical MHD schemes, but
{\em amplified}, so to say, by the special relativistic effects.
These difficulties may be summarized basically in two classes of problems.
The first is concerned with the eigenstructure of the 1-D MHD system, 
which is much more complex than in the fluid case since 
now seven characteristic waves are involved and many different 
degeneracies may occur (depending on the relative orientation 
of the magnetic and velocity vectors). These problems of
{\em non-strict hyperbolicity} can be cured by accurate re-normalizations
of the variables (Brio \& Wu 1988; Roe \& Balsara 1996), to assure
their linear independence, and by introducing additional numerical
viscosity in the degenerate cases. The second aspect is more crucial.
The multidimensional MHD system, in conservative form,
has a specific irreducible structure: the magnetic field
(which is a pseudo-vector) is advanced in time by an antisymmetric
differential operator, a curl, while all other variables are scalars
or vectors advanced in time by a differential operator of divergence form. 
We notice that this basic duality in the conservation laws of the
MHD system is also fully invariant under relativistic coordinate
transformations: the covariant evolution equation for $\vec{B}$
splits into the classical induction equation and in the non-evolutionary
solenoidal $\nabla\cdot\vec{B}=0$ condition.

In numerical schemes where Godunov-type procedures, based on the 
divergence conservation form and on cell-centered variables, 
are also applied to the induction equation, it comes out
that magnetic field components develop unphysical discontinuities 
and numerical monopoles which may grow in time. 
In RMHD this problem can be relevant: when the magnetic field is very
strong, that is when the Alfv\'en velocity approaches the speed of light,
the various eigenvalues collapse one onto the other and it becomes
very hard, from a numerical point of view, to distinguish among different
physical states: thus, even very small errors in the definition of
the magnetic field components, or in the flux derivatives (where the
solenoidal condition is implicitly assumed), often lead to unphysical
states and to code crashing. Therefore, the proper character of the
induction equation and the related preservation of the solenoidal
constraint are fundamental issues in numerical RMHD.

To cope with this class of numerical problems, two families 
of empirical solutions have been proposed: the {\em cleaning}
methods, where the magnetic field components are re-defined at every 
time step (originally proposed by Brackbill \& Barnes, 1980, it requires
the solution of an additional Poisson equation), and the {\em eight-wave}
method (Powell 1994), which modifies the MHD system by adding a new
$\nabla\cdot\vec{B}$ variable, to be advected by the flow like the
other quantities. These methods may in some cases alleviate 
(but not solve) the main difficulties.
A more consistent way to handle this problem is given by the family of
{\em constrained transport} (CT) methods (first introduced by 
Evans \& Hawley, 1988), where the induction equation is
correctly discretized to incorporate the solenoidal constraint as a main
built-in property. Many schemes (Dai \& Woodward 1998; Ryu et al. 1998; 
Balsara \& Spicer 1999b, to cite a few) take advantage of this method, 
but only in a restricted way, since all basic upwind procedures are 
still based on the standard (cell-centered) Godunov-type formalism and
therefore the production of numerical monopoles is not avoided.
A significant further advance has been proposed by KO: after an initial
attempt to extend the eight-wave method to RMHD (failed basically because
of the above mentioned numerical problems), he finally turns to a CT scheme 
where the discretized divergence-free magnetic field components are 
correctly incorporated in the momentum-energy flux functions,
although the upwind fluxes for the induction equations are still
not properly defined, in our opinion.
However, the only astrophysical application of such RMHD code
published so far is the propagation of light relativistic
jets embedded in a purely toroidal magnetic field (Komissarov 1999b),
where the solenoidal condition is actually automatically satisfied
for simple geometrical reasons (the field is bound to remain always 
toroidal), thus this test is not stringent at all for
the solenoidal condition preservation problem.

To date, a fully consistent CT-based upwind scheme
assuring exact $\nabla\cdot\vec{B}=0$ condition has been proposed by
Londrillo \& Del Zanna (2000), LD for brevity. In fact, starting from a
finite volume formulation of the solenoidal condition and 
of the induction equation, as in the original CT method,
general recipes are given to reconstruct (to any order of
accuracy) magnetic field variables at points needed for
flux computations and to formulate approximate Riemann
solvers also for the induction equations in the CT form.
Moreover, as an application, a third order
Essentially Non-Oscillatory (ENO) {\em central}-type
scheme was proposed and numerically validated against various tests.
The so-called central schemes do not make use of time-consuming
and system-dependent spectral decompositions, and linearized Riemann solvers 
are replaced by Lax-Friedrichs type averages over the local Riemann fan. 
In this way, the only characteristic quantities entering the scheme are 
the local fastest velocities, and also the problems related to the various 
degeneracies are thus avoided completely. The price to pay is just 
some additional numerical dissipation at contact and Alfv\'enic 
discontinuities, but the high order reconstruction is often able to 
compensate for these drawbacks. 

The central scheme described in LD was applied to the RHD system
in Del Zanna \& Bucciantini (2002), from now on simply Paper~I,
where the test simulations presented demonstrated the accuracy and 
stability of such scheme, even in highly relativistic situations, 
giving equivalent or better (thanks to its higher order) results 
than those produced by much more elaborate Godunov-type algorithms.
Here the same third order ENO-CT central scheme of LD is extended to the 
RMHD system, thus this paper may be considered as the generalization
of Paper~I to the magnetic case. Therefore, both the structure of
the paper and the formalism used will be the same as in Paper~I, to
which the reader will be often referred, especially for some numerical
scheme details or for test simulations comparisons.
Finally, the CT scheme is extended to generalized orthogonal
curvilinear coordinates in the appendix, thus, the inclusion of
General Relativity effects with a given metric, i.e. the extension to 
GRMHD, may be easily achieved (see the appendix of Koide et al. 1999).

\section{Ideal RMHD equations}

The covariant fluid equations for special relativistic hydrodynamics (RHD)
were given in Paper I and here the same notation will be assumed throughout,
that is all velocities are normalized against the speed of light ($c=1$),
Greek (Latin) indexes indicate four (three) vectors, $g^{\alpha\beta}=
\mathrm{diag}\{-1,1,1,1\}$ is the Minkowski metric tensor (a flat space
is assumed here for ease of presentation, for the extension to any set of 
orthogonal curvilinear coordinates see the appendix), 
and $x^\alpha=(t,x^j)$ is the four vector of space-time coordinates.
The modifications needed to take electromagnetic forces into account are,
like in classical MHD, the inclusion of extra terms in the energy-momentum 
conservation law and a new equation for the magnetic field, to be
derived from Maxwell equations. Our derivation follows that of Anile (1989), 
also described in KO and BA.

Written in terms of the (antisymmetric) electromagnetic tensor 
$F^{\alpha\beta}$ ($F^{0i}=E_i, F^{ij}=B_k$ with $\{i,j,k\}=\{1,2,3\}$
and cyclic permutations) and of its dual $F^{\star\alpha\beta}=
\frac{1}{2}\epsilon^{\alpha\beta\gamma\delta}F_{\gamma\delta}$ 
($F^{\star 0i}=B_i, F^{\star ij}=-E_k$), 
where $\epsilon^{\alpha\beta\gamma\delta}$ is the Levi-Civita 
alternating pseudo-tensor, the covariant Maxwell equations are:
\be
\partial_\alpha F^{\alpha\beta}=-J^\beta,~~~
\partial_\alpha F^{\star\alpha\beta}=0,
\label{maxwell}
\ee
where $J^\alpha$ is the four-current containing the source terms,
constrained by the condition $\partial_\alpha J^\alpha=0$, and 
we have assumed $4\pi\to 1$.
On the other hand, the electromagnetic contribution to the
energy-momentum tensor is
\be
T^{\alpha\beta}_{\rm em}={F^\alpha}_\gamma F^{\beta\gamma}-
\frac{1}{4}g^{\alpha\beta}F_{\gamma\delta}F^{\gamma\delta},
\label{em}
\ee
to be added to the fluid part in the conservation law 
$\partial_\alpha T^{\alpha\beta}=0$.
Finally, we must introduce the covariant relativistic form of 
Ohm's law in the infinite conductivity approximation,
$\vec{E}+\vec{v}\times\vec{B}=0$, that translates into a
condition of vanishing covariant electric field
\be
F^{\alpha\beta}u_\beta=0,
\label{Ohm}
\ee
where $u^\alpha=(\gamma,\gamma v^j)$ is the four-velocity and 
$\gamma\equiv u^0=(1-v^2)^{-1/2}$ is the Lorentz factor.
Note that the other approximation needed to derive the classical MHD
equations, namely to neglect the displacement current, is not imposed
in RMHD, of course, and the result is that the current, to be derived
from the first of Eqs.~(\ref{maxwell}), now depends on the time derivative 
of the electric field too: $\vec{J}=\nabla\times\vec{B}-\partial_t\vec{E}$.

The equations written so far are not easily compared
with their MHD equivalent, due to the presence of the electromagnetic
tensor and of its dual, both containing the electric field. 
However, thanks to Eq.~(\ref{Ohm}), $\vec{E}$ may be substituted
everywhere by defining a magnetic induction four-vector as 
$b^\alpha=F^{\star\alpha\beta}u_\beta$, that allows to write the
electromagnetic tensor in terms of $u^\alpha$ and $b^\alpha$ alone:
$F^{\gamma\delta}=\epsilon^{\alpha\beta\gamma\delta}b_\alpha u_\beta$.
The components of this new four-vector are
\be
b^\alpha=[\gamma (\vec{v}\cdot \vec{B}),
\vec{B}/\gamma+\gamma (\vec{v}\cdot \vec{B})\vec{v}],
\ee
and in the fluid comoving local rest frame we simply have 
$b^\alpha=(0,\vec{B})$.
Note the constraints $u_\alpha b^\alpha=0$ and 
$|u|^2\equiv u_\alpha u^\alpha=-1$, 
so that $|b|^2\equiv b_\alpha b^\alpha>0$ 
and $b^\alpha$ is a space-like vector, with
$|b|^2=B^2/\gamma^2+(\vec{v}\cdot\vec{B})^2$.

Thanks to these definitions, the complete set of RMHD equations becomes:
\be
\partial_\alpha (\rho u^\alpha ) = 0,
\label{mass}
\ee
\be
\partial_\alpha [(w+|b|^2) u^\alpha \! u^\beta - b^\alpha \! b^\beta
+ (p+|b|^2/2)g^{\alpha\beta} ] = 0,
\label{momentum}
\ee
\be
\partial_\alpha (u^\alpha b^\beta - u^\beta b^\alpha)=0,
\label{induction}
\ee
namely the equations of mass conservation, of total energy-momentum 
conservation, and of magnetic induction.
Here $w=e+p$ is the relativistic enthalpy and $e=\rho+p/(\Gamma-1)$ is
the relativistic energy per unit volume for a $\Gamma$-law equation
of state. Notice the analogy with classical MHD equations, easily
obtained by letting $v^2\ll 1$, while RHD equations are recovered
simply by letting $b^\alpha=0$.

\subsection{Evolution equations and the $\nabla\cdot\vec{B}=0$ constraint}

Godunov-type shock-capturing numerical methods developed for classical 
Euler equations apply for any set of hyperbolic conservation laws,
and Paper~I has shown the application to the RHD case.
It is easy to verify that the equations for the fluid variables,
Eqs.~(\ref{mass}) to (\ref{momentum}), retain the usual conservative form:
\be
\frac{\partial \vec{u}}{\partial t} + 
\sum_{i=1}^{3} \frac{\partial \vec{f}^i}{\partial x^i} = 0.
\label{cons_law}
\ee
Here $\vec{u}$ is the vector of conserved variables 
and $\vec{f}^i$ are their corresponding fluxes, along each direction, 
respectively given by
\be
\vec{u}=[\rho u^0,w_\mathrm{tot}u^0u^j-b^0b^j,
w_\mathrm{tot}u^0u^0-b^0b^0-p_\mathrm{tot}]^T,
\label{cons}
\ee
\be
\vec{f}^i\!=[\rho u^i,w_\mathrm{tot}u^iu^j-b^ib^j+
p_\mathrm{tot}\delta^{ij}\!,
w_\mathrm{tot}u^0u^i-b^0b^i]^T\!,
\label{flux}
\ee
where we have defined $w_\mathrm{tot}=w+|b|^2$ and 
$p_\mathrm{tot}=p+|b|^2/2$.

On the other hand, the equation for $b^\alpha$, Eq.~(\ref{induction}), 
splits into two parts, which happen to be exactly the same as 
in classical MHD
(this is not surprising since Maxwell equations are Lorentz invariant).
The spatial component gives the classical induction equation
\be
\frac{\partial\vec{B}}{\partial t}+\nabla\times\vec{E}=0;~~~
\vec{E}=-\vec{v}\times\vec{B},
\label{Bt}
\ee
which is properly the time evolution equation for $\vec{B}$. 
Note that the spatial differential operator is in a {\em curl} form, 
rather than in a {\em divergence} form as Eq.~(\ref{cons_law}). 
This means that the evolution equation of each spatial component 
of $\vec{B}$
has a missing eigen-space, basically due to the antisymmetry of
the electromagnetic tensor in Eq.~(\ref{induction}), as anticipated
in the introduction.
Thus, a total of just three independent magnetic fluxes (the electric field 
vector components, just one in 2-D) are needed for the evolution 
of $\vec{B}$, 
while six independent fluxes were required for the momentum evolution.
The other consequence of the tensor antisymmetric nature is that 
the time component of Eq.~(\ref{induction}) becomes the usual 
MHD solenoidal constraint
\be
\nabla\cdot\vec{B}=0,
\label{sol}
\ee
which is {\em not} an evolutionary equation but a differential constraint
on the spatial derivatives of $\vec{B}$. This constraint is usually regarded 
as just an initial condition, since the form of the induction equation
assures its preservation in time.
Therefore, also numerical schemes must be designed in a way that
the specific divergence-free nature of the magnetic field 
is taken into account as a fundamental constitutive property, otherwise
spurious magnetic monopoles will affect the overall solution and often 
the code stability itself. The CT schemes, and our specific implementation
described in Sect.~3, are the class of numerical schemes based on this 
property.

It is now apparent that Eqs.~(\ref{Bt}) and (\ref{sol}) are substantially
different from the evolutionary conservation laws in Eq.~(\ref{cons_law}).
This fundamental constitutional difference, in both the topology of the
vector $\vec{B}$ field and in its time evolution equation, is
better appreciated by introducing the magnetic vector potential 
$\vec{A}$, defined by $\vec{B}=\nabla\times\vec{A}$,
so that Eq.~(\ref{sol}) is automatically satisfied and Eq.~(\ref{Bt})
takes on the following form (in the Coulomb gauge $\nabla\cdot\vec{A}=0$):
\be
\frac{\partial\vec{A}}{\partial t}+\vec{E}=0.
\label{At}
\ee
For a given velocity field (the so-called kinematic approximation),
Eq.~(\ref{At}) may be regarded as a set of three-dimensional Hamilton-Jacobi 
equations, where the $\vec{E}$ components are functions of the spatial
first derivatives of the $\vec{A}$ components.
Thus, the overall MHD and RMHD systems are actually combinations of 
conservative hyperbolic equations and equations of Hamilton-Jacobi kind.
It is therefore clear that standard numerical upwind schemes developed
for Godunov-type hyperbolic sets of equations {\em cannot} be applied,
and proper upwind expressions for the magnetic flux functions have
to be derived. 

\subsection{Characteristic wave speeds in 1-D RMHD}

The one-dimensional case, say $\partial_y=\partial_z=0$, is, on the other
hand, almost perfectly equivalent to the hydrodynamic case and the overall 
system can be cast in conservative form. In this case Eqs.~(\ref{Bt})
and (\ref{sol}) yield $B_x=\mathrm{const}$ and Eq.~(\ref{cons_law}) becomes
a complete $7\times 7$ system of conservation laws, by just 
adding the $[B_y,B_z]$ variables to (\ref{cons}) and the fluxes 
$[-E_z,E_y]$ to (\ref{flux}). However, the 1-D RMHD system, like its
MHD counterpart, is not {\em strictly} hyperbolic, in the sense that two or 
more eigenvalues may coincide in some degenerate cases, depending on the
angle between the direction of propagation and the local magnetic field.

The characteristic structure of this system was first studied by
Anile and Pennisi (1987; see also Anile, 1989), who derived the
eigenvalues and eigenvectors of the associated Jacobian $\partial\vec{f}/
\partial\vec{u}$ by using the covariant notation. All the particular
degeneracies were also taken into account.

However, since in our numerical scheme the detailed characteristic 
eigenstructure is not required, while only the two speeds 
at the local Riemann fan boundaries need to computed, 
here we just report the expressions for the eigenvalues, 
in the form shown in the appendix of KO. These are one entropy wave
\be
\lambda_0=v_x,
\ee
two Alfv\'en waves
\be
\lambda_\mathrm{A}^\pm=\frac{u^x\pm \tilde{b}^x}{u^0\pm \tilde{b}^0},
\ee
and four magneto-sonic waves (two fast and two slow waves), that unfortunately
do not have a simple analytical expression and must be derived from 
the nonlinear quartic equation
\begin{displaymath}
(1-\varepsilon^2)(u^0\lambda-u^x)^4
\end{displaymath}
\be
~~~+(1-\lambda^2)[(c_s^2(\tilde{b}^0\lambda-\tilde{b}^x)^2-
\varepsilon^2(u^0\lambda-u^x)^2]=0,
\label{ms}
\ee
where $c_s^2=\Gamma p/w$ is the sound speed squared, 
$\tilde{b}^\alpha=b^\alpha/\sqrt{w_\mathrm{tot}}$ 
($|\tilde{b}|^2=\tilde{b}_\alpha\tilde{b}^\alpha=|b|^2/w_\mathrm{tot})$, 
and $\varepsilon^2=c_s^2+|\tilde{b}|^2-c_s^2|\tilde{b}|^2$. Note that the
ordering of MHD characteristic speeds and the various degeneracies are 
preserved in the relativistic case (Anile 1989), although the symmetry
between each $\lambda^\pm$ couple of waves is lost, due to relativistic
aberration effects.

Several numerical algorithms may be employed to solve Eq.~(\ref{ms}):
Newton's root finding technique, applied in the proper interval
for each characteristic speed, Laguerre's method for polynomials, 
involving complex arithmetics, or eigenvalues finding routines, based
on the associated {\em upper Hessenberg} matrix. We have tested all
these numerical methods by using the {\em Numerical Recipes} 
(Press et al. 1986) appropriate routines, which are {\em rtsafe},
{\em zroots}, and {\em hqr}, respectively, under a wide range of
conditions, including various degenerate cases and ultra-relativistic 
speeds, temperatures or magnetic fields.
However, in the code we have decided to adopt the analytical approach, 
described for example in Abramowitz \& Stegun (1965), which requires in turn
the analytical solution of a cubic and of two quadratic algebraic equations.
We have found that this algorithm gives results comparable to Laguerre's or
the matrix methods, the most robust and precise ones, and it is much less 
computationally expensive. On the other hand, Newton's iterative method,
which is the fastest in normal conditions, was found to fail in some nearly 
degenerate cases.

\subsection{Primitive variables}

In order to compute fluxes, the vector of primitive fluid variables 
$\vec{v}=[\rho,v^j,p]^T$ have to be recovered from the conservative ones 
$\vec{u}=[D,Q^j,E]^T$, defined in Eq.~(\ref{cons}), at the beginning of 
every numerical time step (please notice that in the present sub-section 
$E$ will indicate the total energy, which has nothing to share with the 
electric field $\vec{E}$). Like for RHD codes, this procedure must be 
carried out by some iterative root-finding routine. 
In the magnetic case this process is even more difficult, in spite of
the fact that $\vec{B}$ can be considered as given (its components are 
both primitive and conserved variables).
In BA the full $5\times 5$ system is solved by inverting the
$\vec{u}(\vec{v})=0$ set of nonlinear equations as they stand,
providing also all the partial derivatives needed. 
However, we have verified that this process is neither efficient nor
stable when relativistic effects are strong. 
In Koide et al. (1996) and KO the system to solve was reduced down 
to a couple nonlinear equations, while here we manage to derive just 
{\em one} nonlinear equation to be solved iteratively, with
obvious improvements both in terms of speed and precision.

The first step is to use the definitions of $u^\alpha$ and $b^\alpha$
to write the known conservative variables $\vec{u}=[D,Q^j,E]^T$ in
terms of the primitive variables. The vector $\vec{Q}$ becomes
\be
\vec{Q}=(W+B^2)\vec{v}-(\vec{v}\cdot\vec{B})\vec{B},
\label{Q}
\ee
and by taking the projection along $\vec{B}$ we find the important
relation $S\equiv(\vec{Q}\cdot\vec{B})=W(\vec{v}\cdot\vec{B})$, where
like in Paper I we have used $W=w\gamma^2$, $w=\rho+\Gamma_1p$, 
and $\Gamma_1=\Gamma/(\Gamma-1)$.

A $2\times 2$ system of nonlinear equations is then derived by taking the 
square of Eq.~(\ref{Q}) and by using the equation for the total energy $E$:
\be
W^2v^2+(2\,W+B^2)B^2v_\perp^2-Q^2=0,
\label{Q2}
\ee
\be
W-p+\frac{1}{2}B^2+\frac{1}{2}B^2v_\perp^2-E=0,
\label{E}
\ee
where $B^2v_\perp^2\equiv B^2v^2-(\vec{v}\cdot\vec{B})^2=B^2v^2-S^2/W^2$.
If we then use the relations
\be
\rho=D\sqrt{1-v^2},~~~p=[(1-v^2)W-\rho]/\Gamma_1,
\label{prho}
\ee
it comes out that
all quantities appearing in the system are written in terms of the
two unknowns $v^2$ (or equivalently $\gamma$) and $W$. Once these
variables are found numerically, primitive quantities will be easily
derived through Eqs.~(\ref{prho}) and by inverting Eq.~(\ref{Q}), that is
\be
\vec{v}=\frac{1}{W+B^2}\left(\vec{Q}+\frac{S}{W}\vec{B}\right).
\ee

In order to bring the system down to just a single nonlinear equation
(to be solved numerically, for example by Newton's iterative method), 
we found it useful to define $B^2v_\perp^2=T^2/(W+B^2)^2$ in Eqs.~(\ref{Q2}) 
and (\ref{E}), where $T^2\equiv B^2Q^2-S^2$ is a new, but given, parameter.
Then we write Eq.~(\ref{E}) as a third order algebraic equation 
for $W$ with coefficients that depend on $v^2$ alone
\be
\left[\left(1\!-\!\frac{1\!-\!v^2}{\Gamma_1}\right)\!W\!-
\!E+\!\frac{\rho}{\Gamma_1}\!+\!\frac{B^2}{2}
\!\right]\!(W\!+\!B^2)^2\!+\!\frac{T^2}{2}\!=\!0,
\label{W}
\ee
which can be solved analytically (again, see Abramowitz \& Stegun 1965).
Note that the cubic polynomial on the left hand side has
a positive local maximum in $W=-B^2$. Thus, since we know that at
least one root must be positive, all the three roots of Eq.~(\ref{W}) 
are actually bound to be real, and we have verified that the largest one
always yields the correct result.

The function $W(\xi)$, with $\xi=v^2$, is thus available together
with its derivative $W^\prime(\xi)$, so the final step is to apply
Newton's method to find the root of ${\cal F}(\xi)=0$, where
\be
{\cal F}(\xi)=W^2\xi+(2\,W+B^2)\frac{T^2}{(W+B^2)^2}-Q^2,
\label{F}
\ee
and
\be
{\cal F}^\prime(\xi)=W^2+2\,WW^\prime\left[\xi-\frac{T^2}{(W+B^2)^3}\right].
\label{F1}
\ee
The numerical routine actually employed in the code is {\em rtsafe}
(Press et al. 1986), which applied with an accuracy of $10^{-6}$ 
in the range $\xi_\mathrm{min}=0$ and $\xi_\mathrm{max}=1-10^{-6}$ 
($\gamma_\mathrm{max}=1000$), typically converges in 5--10 iterations 
for any set of conservative variables and magnetic field components 
that actually admits a solution.

The technique described above appears to be 
extremely efficient and, above all, robust; we therefore recommend
its use in all RMHD shock-capturing codes, whatever the numerical scheme 
actually employed.

\section{The finite-difference ENO-CT central scheme}

In the present section the third order ENO-CT scheme described in LD
will be adapted and applied to the RMHD equations derived above.
The ENO-CT scheme employs a finite-difference discretization framework, 
and uses {\em Convex} ENO (CENO) reconstruction procedures
to get high order non-oscillatory point values of primitive variables needed
to compute numerical flux derivatives. The upwind procedures in numerical
fluxes are based on approximate Riemann solvers of the Lax-Friedrichs 
type, as in other central schemes. In particular, here we adopt a flux 
formula based on {\em two} local characteristic speeds, rather than just 
one as in the code discussed in LD. The scheme will be here presented in
the semi-discrete formalism, that is time dependency is implicitly
assumed for all spatially discretized quantities. The evolution
equations are then integrated in time by applying a third order
TVD Runge-Kutta algorithm (Shu \& Osher 1988), with a time-step
inversely proportional to the largest (in absolute value) of the 
characteristic speeds (the magneto-sonic velocities defined in Sect.~2.2) 
present in the domain and subject to the CFL condition. 
The reader is referred to both LD and Paper~I
for further numerical references and comments. In any case, see Shu (1997)
for a general overview of ENO schemes, Liu \& Osher (1998) for the original
formulation of the CENO central scheme, and Kurganov et al. (2001) for
the introduction of two-speed averaged Riemann solvers in high order
central schemes for both hyperbolic and Hamilton-Jacobi equations. 
 
\subsection{Discretization of fluid and magnetic variables}

Given a Cartesian uniform 3-D mesh of $N_x\times N_y\times N_z$ cells
(volumes), with sizes $\Delta x$, $\Delta y$, and $\Delta z$, 
let us indicate with $P_{i,j,k}\equiv (x_i,y_j,z_k)$ the cell centers 
and with $P_{i+1/2,j,k}$ the points centered on intercell surfaces 
(along $x$ in this case).
Classical Godunov-type schemes are usually formulated in a finite-volume 
(FV) framework, where state variables are advanced in time as cell averaged 
values. By applying Gauss' theorem to Eq.~(\ref{cons_law}), the
time evolution equation for the vector of {\em fluid} variables becomes
\be
\frac{\mathrm{d}}{\mathrm{d}t}\bar{\vec{u}}=-\sum_{i=1}^{3}
\frac{\Delta_i\bar{\vec{f}}^i}{\Delta x^i},~\mathrm{at}~P_{i,j,k},
\label{dudt_fv}
\ee
where, for each scalar variable, $\bar{u}\equiv\bar{u}_{i,j,k}$ 
is the FV cell averaged discretization of $u(\vec{x})$ at $P_{i,j,k}$ .
Flux derivatives are given in conservative form as simple two-point 
differences of the numerical fluxes $\bar{f}^i$. These fluxes are defined
as intercell surface averages and are stored on surface centers,
each in the direction corresponding to the $i$ component, with $i=1,2,3$.
For example, $\bar{f}=\bar{f}^x$ is located on $P_{i+1/2,j,k}$ points, 
and the $\Delta_x$ operator, centered in $P_{i,j,k}$, is defined as
\be
[\Delta_x\bar{f}]_{i,j,k}=\bar{f}_{i+1/2,j,k}-\bar{f}_{i-1/2,j,k}.
\ee
Similar expressions hold in the other directions.

A different strategy is needed to reconstruct the magnetic field 
variables. The induction equation Eq.~(\ref{Bt}) is in curl form,
thus the correct procedure for discretization in the FV framework
is the application of Stokes' theorem. The $x$ component gives
\be
\frac{\mathrm{d}}{\mathrm{d}t}\bar{B}_x=
-\frac{\Delta_y\bar{E}_z}{\Delta y}+\frac{\Delta_z\bar{E}_y}{\Delta z},
~\mathrm{at}~P_{i+1/2,j,k},
\label{Bt_num}
\ee
where $\bar{B}_x$ is discretized as  {\em surface} average and located 
at $P_{i+1/2,j,k}$ intercell points, while for example $\bar{E}_z$
is a {\em line} average and is located at $P_{i+1/2,j+1/2,k}$ volume
{\em edge} points (cell corners in 2-D). 
Similar expressions are defined for the $y$ and $z$ components.
Thanks to the above discretization, it is straightforward to prove
that the numerical solenoidal condition will be algebraically satisfied 
at all times (if satisfied at $t=0$):
\be
\frac{\mathrm{d}}{\mathrm{d}t}\left(
\frac{\Delta_x\bar{B}_x}{\Delta x}+
\frac{\Delta_y\bar{B}_y}{\Delta y}+
\frac{\Delta_z\bar{B}_z}{\Delta z}
\right)\equiv 0,~\mathrm{at}~P_{i,j,k}.
\label{divb}
\ee
This is the fundamental property of the CT method and relies
on the definition of the staggered field components
\be
[\bar{B}_x]_{i+1/2,j,k},\,[\bar{B}_y]_{i,j+1/2,k},\,[\bar{B}_z]_{i,j,k+1/2}
\label{primary}
\ee
as primary data.
It is important to notice that these data contain essential informations
not only for the discretization of the corresponding variables at cell 
centers, but also for the definition of longitudinal derivatives at the 
same points, since two values per cell are available for each component.

A first consequence is that a {\em continuous} numerical vector potential 
$\bar{\vec{A}}$ can be derived in a unique way by inverting the discretized 
form of the $\nabla\times\vec{A}=\vec{B}$ relation. 
By applying Stokes' theorem once more it is easy to verify that
these data must be defined as line averages along the longitudinal
direction and stored at the same locations as the corresponding electric 
field $\bar{\vec{E}}$ components. 
On the other hand, if the $\bar{\vec{A}}$ components are used 
as primary data, then the time evolution discretized equations
in Eq.~(\ref{Bt_num}) must be replaced by 
\be
\frac{\mathrm{d}}{\mathrm{d}t}\bar{A}_x=-\bar{E}_x
~\mathrm{at}~P_{i,j+1/2,k+1/2},
\label{At_fv}
\ee
and similarly for the other components. The field components in 
Eq.~(\ref{primary}) are then derived as
\be
\bar{B}_x=\frac{\Delta_y\bar{A}_z}{\Delta_y}-
\frac{\Delta_z\bar{A}_y}{\Delta_z}~\mathrm{at}~P_{i+1/2,j,k},
\ee
and similarly for the $y$ and $z$ components. In the CT framework, 
the two choices are perfectly equivalent, and the solenoidal constraint
Eq.~(\ref{divb}) is still clearly satisfied exactly.

A second main property is given by the continuity condition of the 
face averaged components (i.e. the numerical magnetic fluxes)
in Eq.~(\ref{primary}). For example, the numerical function
\be
\bar{B}_x(x)=\frac{1}{S}\int_{S}B_x(\vec{x})\,\mathrm{d}S,
\label{stag}
\ee
is a continuous function of $x$, where $S$ is a cell section normal
to the $x$ direction. A simple proof is given by integrating
$\nabla\cdot\vec{B}=0$ on a volume including $S$ and tending to it
by letting $\Delta x\to 0$ (see LD). Therefore, at points of discontinuity
(intercell surfaces) staggered data are well defined as point values 
in the corresponding longitudinal direction, having a single-state 
numerical representation there (this fundamental property will be fully
appreciated later on for upwind calculations) and showing at most 
discontinuous first derivatives along the corresponding coordinate.

In our ENO-CT scheme, the FV discretization is replaced by a
{\em finite-difference} (FD) discretization based on point values, 
which is more efficient in the multi-dimensional case and 
allows to use just 1-D interpolation routines.
The primary data actually employed in our scheme are thus the
point-valued $\vec{u}$ fluid variables, defined at cell centers
$P_{i,j,k}$, and the point-valued potential vector $\vec{A}$ components,
located at cell edges exactly as in the FV approach.
The time evolution equations are thus, in the FD CT scheme
\be
\frac{\mathrm{d}}{\mathrm{d}t}\vec{u}=-\sum_{i=1}^{3}
\frac{\Delta_i\hat{\vec{f}}^i}{\Delta x^i},~\mathrm{at}~P_{i,j,k},
\label{dudt_fd}
\ee
and
\begin{eqnarray}
(\mathrm{d}/\mathrm{d}t)A_x & = & -E_x,
~\mathrm{at}~P_{i,j+1/2,k+1/2}, \nonumber \\
(\mathrm{d}/\mathrm{d}t)A_y & = & -E_y,
~\mathrm{at}~P_{i+1/2,j,k+1/2}, \label{At_fd} \\
(\mathrm{d}/\mathrm{d}t)A_z & = & -E_z,
~\mathrm{at}~P_{i+1/2,j+1/2,k}. \nonumber 
\end{eqnarray}
where now the numerical fluid flux functions $\hat{\vec{f}}^i$ are
such that their volume average approximate the FV fluxes $\bar{\vec{f}}^i$
to the given order of accuracy, while electric fields in the FD formalism
are simply point-valued numerical functions.
Eqs.~(\ref{dudt_fd}) and (\ref{At_fd}) are thus the time evolution
equations that are integrated by the Runge-Kutta time-stepping
algorithm. The reconstruction procedures and the upwind formulae
to define numerical fluxes $\hat{\vec{f}}^i$ and $\vec{E}$, in the
respective locations, are given in the following sub-sections.

\subsection{Reconstruction procedures}

At this preliminary level of analysis, everything is exact. 
Approximations come into play only in the reconstruction procedures, 
when point values needed for flux computations are recovered from 
primary numerical data to some accuracy level.
At a second order approximation, the procedures are straightforward, 
since the two discretization approaches, FV and FD, coincide.
At higher order of accuracy, on the other hand, specific procedures
must be defined.

The first step is to derive magnetic field components from vector
potential data. This is done in our scheme at the beginning of
each time-stepping sub-cycle. To third order accuracy, we first define
\begin{eqnarray}
\hat{A}_x & = & [1-\gamma_1 {\cal D}_y^{(2)}-\gamma_1 {\cal D}_z^{(2)}]A_x,
~\mathrm{at}~P_{i,j+1/2,k+1/2}, \nonumber \\
\hat{A}_y & = & [1-\gamma_1 {\cal D}_z^{(2)}-\gamma_1 {\cal D}_x^{(2)}]A_y,
~\mathrm{at}~P_{i+1/2,j,k+1/2}, \\
\hat{A}_z & = & [1-\gamma_1 {\cal D}_x^{(2)}-\gamma_1 {\cal D}_y^{(2)}]A_z,
~\mathrm{at}~P_{i+1/2,j+1/2,k}, \nonumber 
\end{eqnarray}
where $\gamma_1=1/24$ and ${\cal D}^{(2)}$ is the non-oscillatory
numerical second derivative defined in Paper~I. 
Then the divergence-free magnetic field components are derived directly
from the high order approximation of $\vec{B}=\nabla\times\vec{A}$, which
gives
\begin{eqnarray}
\hat{B}_x & = & (\Delta_y\hat{A}_z)/\Delta y-(\Delta_z\hat{A}_y)/\Delta z,
~\mathrm{at}~P_{i+1/2,j,k}, \nonumber \\
\hat{B}_y & = & (\Delta_z\hat{A}_x)/\Delta z-(\Delta_x\hat{A}_z)/\Delta x, 
~\mathrm{at}~P_{i,j+1/2,k}, \label{B_hat} \\
\hat{B}_z & = &(\Delta_x\hat{A}_y)/\Delta x-(\Delta_y\hat{A}_x)/\Delta y,
~\mathrm{at}~P_{i,j,k+1/2}. \nonumber
\end{eqnarray}
These new staggered field components clearly satisfy, at all times $t$:
\be
\frac{\Delta_x\hat{B}_x}{\Delta x}+
\frac{\Delta_y\hat{B}_y}{\Delta y}+
\frac{\Delta_z\hat{B}_z}{\Delta z}
\equiv 0,~\mathrm{at}~P_{i,j,k},
\label{divb_fd}
\ee
which is the point-value equivalent expression of Eq.~(\ref{divb}). 
Thus, in the FD version of the CT scheme, the fundamental divergence-free
magnetic components are those defined in Eq.~(\ref{B_hat}), 
whose divided differences directly give high order approximations 
of the longitudinal derivative.

Having assured approximated first derivatives satisfying exact
divergence-free relations, it is now possible to reconstruct the
corresponding point values in the longitudinal coordinate, which
will be needed for the definition of numerical fluxes, in a way
to maintain the value of the first derivative.
Since only one-dimensional operators, 
in turn, are now required, we denote by $B$ the unknown point-value
numerical data and by $\hat{B}$ the data derived just above, where
both sets are located at the same intercell points.
By definition, to third order approximation we have
\be
[1-\gamma_1 {\cal D}^{(2)}] B = \hat{B}.
\ee
For a given set of values $\hat{B}$, one has then to invert the
${\cal D}^{(2)}$ operator, which is not based on a fixed stencil of data
and the resulting matrix appears then to be highly non-linear. However,
the operator on the left hand side can be inverted in an explict
way by using the Taylor expansion
$[1-\gamma_1 {\cal D}^{(2)}]^{-1}\simeq 1+
[\gamma_1 {\cal D}^{(2)}]+[\gamma_1 {\cal D}^{(2)}]^2+\dots$,
so that the component $B$ may be approximated
by a finite order iteration as
\be
B^{(n)}=\hat{B}+[\gamma_1 {\cal D}^{(2)}] B^{(n-1)};~~~B^{(0)}=\hat{B}.
\label{iter}
\ee
It is essential to notice that the approximation order of the $B$ point
values does not increases with the iteration number $n$, being
always of the third order of the base scheme. What changes is the
residual error in the related solenoidal condition:
usually $n=5$ iterations are enough to assure the preservation
of the longitudinal derivative value, so that Eq.~(\ref{divb_fd})
remains exact within machine accuracy in the computation of flux 
derivatives too, thus avoiding spurious monopoles terms in the 
dynamical equations.

A final interpolation step is needed to define point-value magnetic field 
$\vec{B}$ components at cell centers $P_{i,j,k}$, where fluid conservative
variables $\vec{u}$ are stored. Thus, at the beginning of each time
sub-cycle, point-value primitive variables $\vec{v}=[\rho,v^j,p]^T$
can be recovered as described in Sect.~2.3.

In order to use these variables in the definition of numerical fluxes,
a reconstruction step is required along each direction, to provide
point-value data at intercell points.
For multi-dimensional calculations and for higher than second order
schemes, like in our case, the reconstruction routines are
one-dimensional only in the FD framework, which is then to be preferred.
Moreover, the reason for reconstructing the primitive variables
is also apparent: if reconstruction were applied to conservative
variables, then the time-consuming (in RMHD) algorithm of Sect.~2.3
would be needed at intercell points for each direction.
The reconstruction routines employed in our code are the Convex ENO
procedures described in Paper~I, with a choice of two slope limiters
(MinMod and Monotonized Centered) to prevent unwanted oscillations
and to preserve monotonicity.

Reconstruction procedures at intercell locations give
two-state, left (L) and right (R), reconstructed variables, 
depending on the stencil used in the definition of the (quadratic) 
interpolation polynomial. However, not all the eight variables retain
such two-state representation, since we know from Eq.~(\ref{stag})
that the longitudinal field component along the direction
of flux differentiation must be continuous at corresponding
intercell locations, and this property is preserved also for point-value 
staggered $\vec{B}$ components defined through Eq.~(\ref{iter}).
Therefore, the longitudinal field is not reconstructed as the other
seven Godunov variables, and the single-state value provided
by Eq.~(\ref{iter}) is assumed in flux calculations.
Notice that this is the crucial point discussed in the introduction: 
the use of the divergence-free field components in numerical fluxes 
permits to avoid the onset of spurious monopoles in the computation
of the right-hand side of Eq.~(\ref{dudt_fd}).

Finally, point-value numerical fluxes $\vec{f}$ defined at intercell
locations (the proper upwind procedures to define them will be
discussed in the following sub-section) have to be further transformed
in the corresponding $\hat{\vec{f}}$ fluxes defined in Eq.~(\ref{dudt_fd}).
This step is required to approximate flux derivatives to higher than
second order. To third order accuracy we have, as usual
\be
\hat{\vec{f}}=[1-\gamma_1 {\cal D}^{(2)}] \vec{f},
\ee
whereas this final step is not needed for electric fields, which
are defined in Eqs.~(\ref{At_fd}) as point-value data.

\subsection{Central-upwind numerical fluxes}

In Lax-Friedrichs central-type schemes, two couples of characteristic 
velocities $\lambda_\pm^\mathrm{L}$ and $\lambda_\pm^\mathrm{R}$
are first defined at intercell locations. These velocities,
which are the fast magneto-sonic speeds in RMHD, are those at the boundaries 
of the two local Riemann fans (one fan for each $L$ or $R$ state),
and are derived by using the procedure described in Sect.~2.2.
Then a local averaged Riemann problem is solved by using either the
two-speed HLL (from Harten, Lax, and van Leer) or the single-speed LLF
(local Lax-Friedrichs) flux formulae:
\be
\vec{f}^{\rm HLL}=
\frac{\alpha^+\vec{f}^\mathrm{L}+\alpha^-\vec{f}^\mathrm{R}
-\alpha^+\alpha^-(\vec{u}^\mathrm{R}-\vec{u}^\mathrm{L})}
{\alpha^++\alpha^-},
\label{hll}
\ee
\be
\vec{f}^{\rm LLF}=\frac{1}{2}
[\vec{f}^\mathrm{L}+\vec{f}^\mathrm{R}-\alpha
(\vec{u}^\mathrm{R}-\vec{u}^\mathrm{L})],
\label{llf}
\ee
where
\be
\alpha^\pm=\mathrm{max}\{0,\pm\lambda_\pm^\mathrm{L},
\pm\lambda_\pm^\mathrm{R}\};~~~
\alpha=\mathrm{max}\{\alpha^+,\alpha^-\},
\label{alpha}
\ee
where basically all the other intermediate states are averaged out.
Notice that when the local Riemann fan is symmetric, then $\alpha^+=
\alpha^-=\alpha$ and the two fluxes coincide, whereas, when both
fast magneto-sonic speeds have the same sign one of the $\alpha^\pm$ is 
zero and the HLL flux becomes a pure upwind flux, either 
$\vec{f}^\mathrm{L}$ or $\vec{f}^\mathrm{R}$. This is why the
HLL scheme described above was defined in Kurganov et al. (2001)
as {\em central-upwind}.

The upwind states for the numerical $\vec{E}$ flux functions of the induction
equation are defined by the same averaged Riemann solver, now to be applied
to flux functions having a base four-state structure.
This different structure arises because at cell edges, 
where the electric fields must be defined, {\em two} surfaces
of discontinuity intersect, and modes of Riemann fans coming from 
different directions overlap there.
A proper expression that extends Eq.~(\ref{hll}) and (\ref{llf}) 
to the induction equation fluxes is derived by taking advantage of the 
analytical and numerical experience developed for the Hamilton-Jacobi 
equations. For ease of presentation, only the $z$ component of the electric 
field, $E_z=-(v_xB_y-v_yB_x)$ (that usually needed in 2-D simulations), 
will be treated here, while the $x$ and $y$ components are easily obtained 
by cyclic permutations of the indexes.

Let us indicate with double upper indexes these four states, where
the first refers to upwinding along $x$ and the second along $y$,
obtained by reconstructing the required primitive variables at the 
edge point $P_{i+1/2,j+1/2,k}$ by applying a sequence of two 
independent one-dimensional reconstruction routines.
For each component of the magnetic field just one reconstruction in
the transverse direction is actually required, of course.
The proposed HLL and LLF upwind formulae for the $E_z$ flux function 
are given respectively by
\begin{eqnarray}
\lefteqn{E_z^\mathrm{HLL} =\frac{
\alpha_x^+\alpha_y^+E_z^\mathrm{LL}\!+\!
\alpha_x^+\alpha_y^-E_z^\mathrm{LR}\!+\!
\alpha_x^-\alpha_y^+E_z^\mathrm{RL}\!+\!
\alpha_x^-\alpha_y^-E_z^\mathrm{RR}
}{(\alpha_x^++\alpha_x^-)(\alpha_y^++\alpha_y^-)}} \nonumber \\
& & +\frac{\alpha_x^+\alpha_x^-}{\alpha_x^++\alpha_x^-}
(B_y^\mathrm{R}-B_y^\mathrm{L})
-\frac{\alpha_y^+\alpha_y^-}{\alpha_y^++\alpha_y^-}
(B_x^\mathrm{R}-B_x^\mathrm{L}), \label{hll_e} \\
\lefteqn{E_z^\mathrm{LLF}= \frac{1}{4}
(E_z^\mathrm{LL}\!+\!E_z^\mathrm{LR}\!+
\!E_z^\mathrm{RL}\!+\!E_z^\mathrm{RR})}
\nonumber \\
& & +\frac{1}{2}\alpha_x(B_y^\mathrm{R}-B_y^\mathrm{L})
-\frac{1}{2}\alpha_y(B_x^\mathrm{R}-B_x^\mathrm{L}). \label{llf_e}
\end{eqnarray}
The $\alpha_x^\pm$ and $\alpha_y^\pm$ at $P_{i+1/2,j+1/2,k}$ required above
should be calculated by taking the maximum characteristic speed (in absolute
value) among the four reconstructed states, whereas for sake of efficiency
we actually consider the maximum over the two neighboring inter-cell points, 
where these speeds had been already calculated for fluid fluxes.
As usual, the LLF numerical flux is obtained from the HLL one by letting
$\alpha_x^+=\alpha_x^-=\alpha_x$ and $\alpha_y^+=\alpha_y^-=\alpha_y$.

Note that for a pure 2-D case $\partial_xA_z=-B_y$ and $\partial_yA_z=+B_x$,
thus for a given velocity field the induction equation simply becomes
the Hamilton-Jacobi equation $(\mathrm{d}/\mathrm{d}t)A_z=
-E_z(\partial_xA_z,\partial_yA_z)$ and our upwind formulae correctly
match in this case with those given in Kurganov et al. (2001), where
the HLL central-upwind scheme was applied to this kind of equations.
Moreover, notice that for discontinuity surfaces coincident with one
of the inter-cell boundaries, that is for 1-D situations, Eqs.~(\ref{hll_e})
and (\ref{llf_e}) reduce respectively to Eqs.~(\ref{hll}) and (\ref{llf}),
as it should be. Thus, for 2-D or 1-D calculations, our 3-D ENO-CT scheme
automatically treats the magnetic field components which does not require
a Hamilton-Jacobi formulation in the usual Godunov-type approach.

The fact that the magnetic numerical fluxes for the induction equation
must use upwind formulae based on four-state quantities seems to have been
overlooked by the other authors, who generally just {\em interpolate} 
the $x$ and $y$ 1-D single-state upwind fluxes already
calculated at intercell points to the cell corner where $E_z$ 
must be defined (we specialize here to a 2-D situation). 
However, this procedure is clearly incorrect, because at cell corners
only $B_x$ and $B_y$ have a two-state representation, whereas $v_x$
and $v_y$ retain the complete four-state representation.

\section{Numerical results}

\begin{figure*}
\centerline{\resizebox{.85\hsize}{!}{\includegraphics{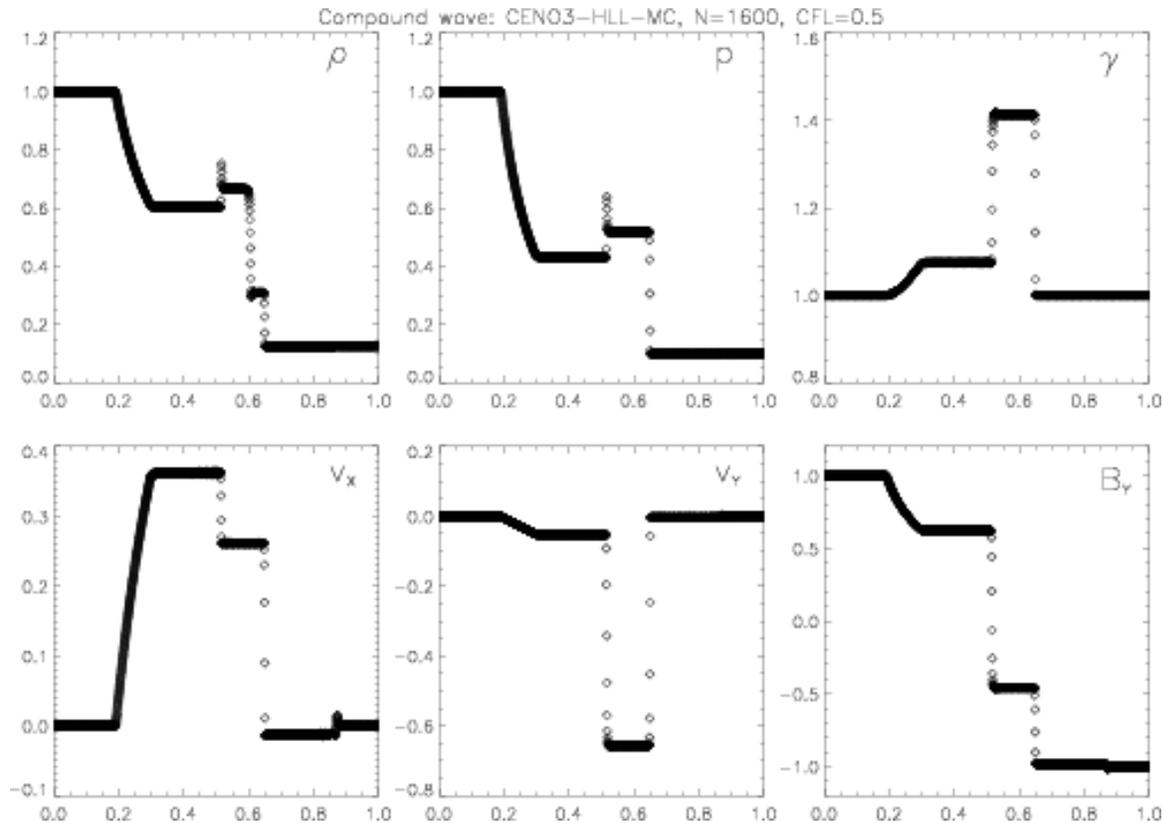}}}
\caption{The relativistic analog of Brio \& Wu (1988) MHD test problem
involving a compound wave. If compared to BA, our left moving compound 
shock and right moving slow shock are better resolved. Here the base
scheme CENO3-HLL-MC is used, with $N=1600$ grid points to compare with
BA and Courant number CFL$=0.5$.}
\end{figure*}

\begin{figure*}
\centerline{\resizebox{.85\hsize}{!}{\includegraphics{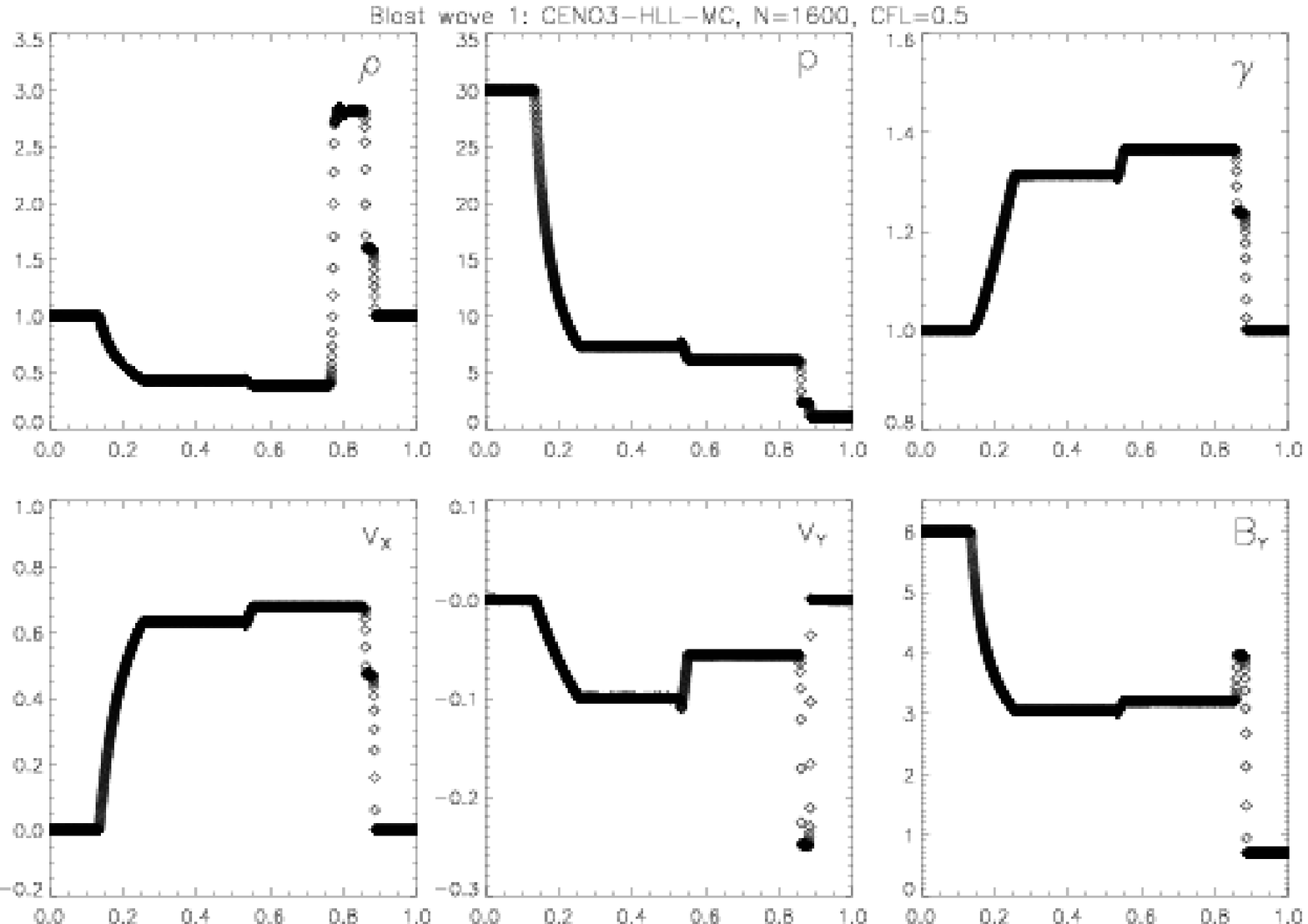}}}
\vspace*{1cm}
\centerline{\resizebox{.85\hsize}{!}{\includegraphics{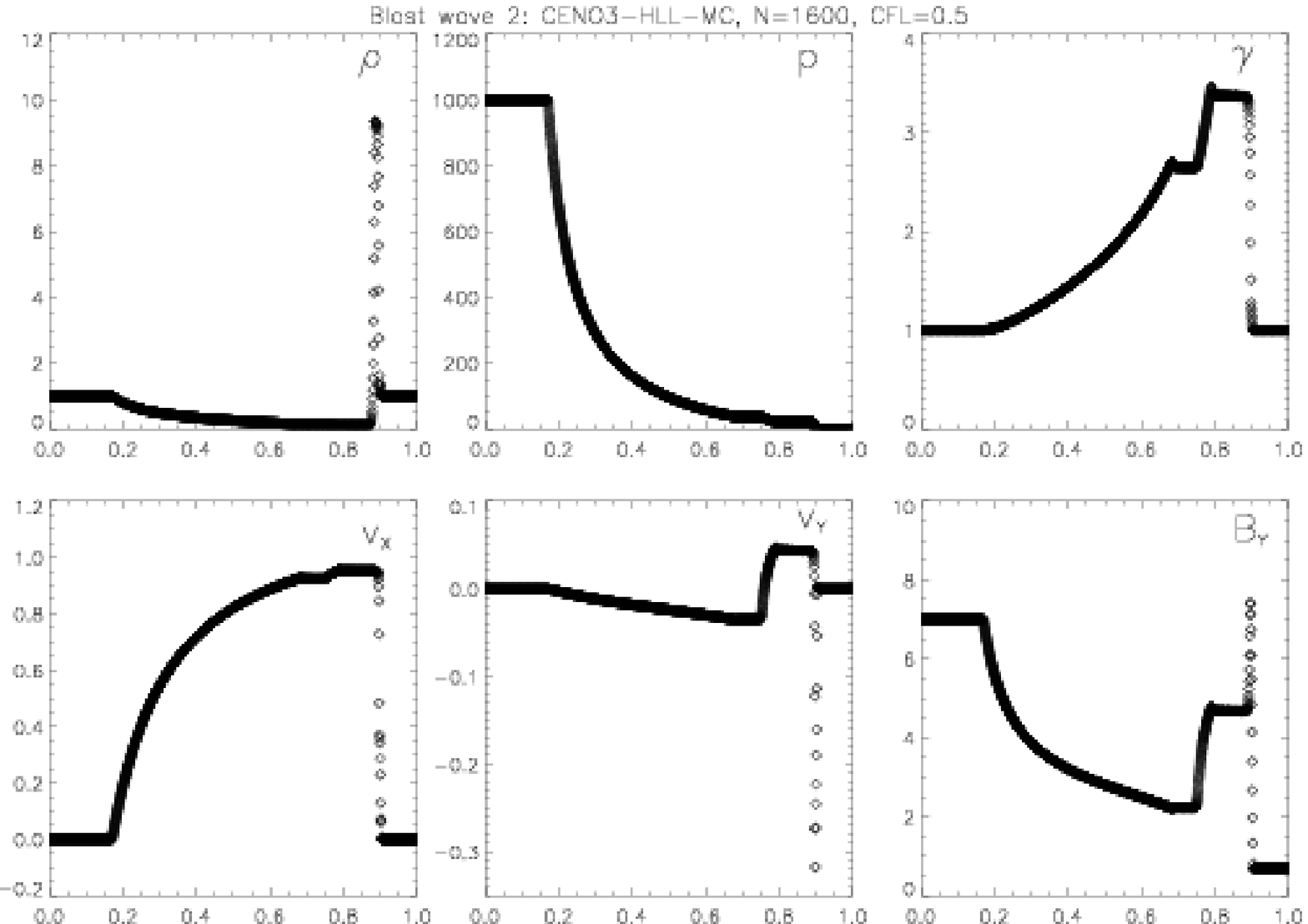}}}
\caption{A couple of relativistic, magnetized blast waves. The first
(upper panel) has a moderate initial pressure jump 
($p^\mathrm{L}/p^\mathrm{R}=30$),
whereas the second (bottom panel) has a much stronger jump 
($p^\mathrm{L}/p^\mathrm{R}=10^4$),
producing a very narrow density peak and a Lorentz factor of 
$\gamma\simeq 3.4$. Numerical settings are the same as in Fig.~1.}
\end{figure*}

\begin{figure*}
\centerline{\resizebox{.85\hsize}{!}{\includegraphics{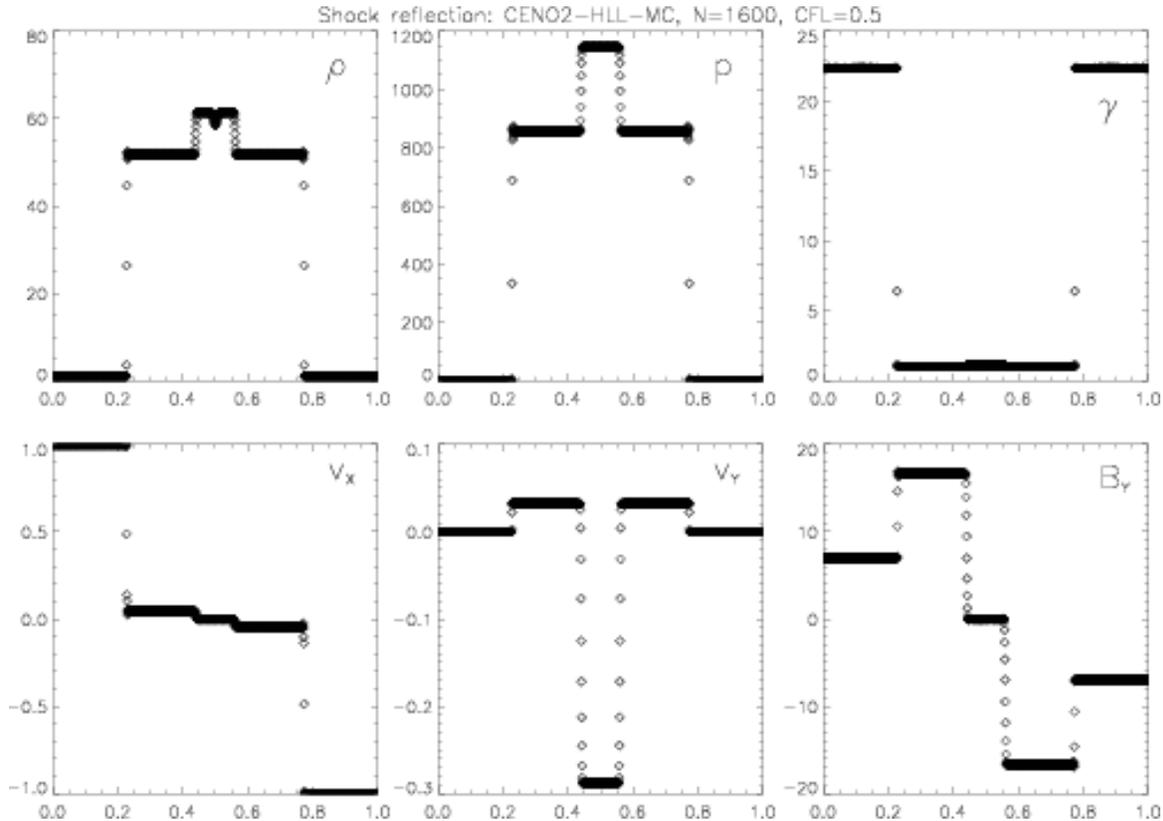}}}
\caption{The relativistic MHD shock reflection problem, with
$\gamma\simeq 22.4$ flows colliding at $x=0.5$.
This test is crucial for two reasons: the post-shock oscillations, here
damped by reconstructing at second order, and the {\em wall heating} 
problem, that appears to be quite reduced by the use of HLL.
}
\end{figure*}

\begin{figure*}
\centerline{\resizebox{.85\hsize}{!}{\includegraphics{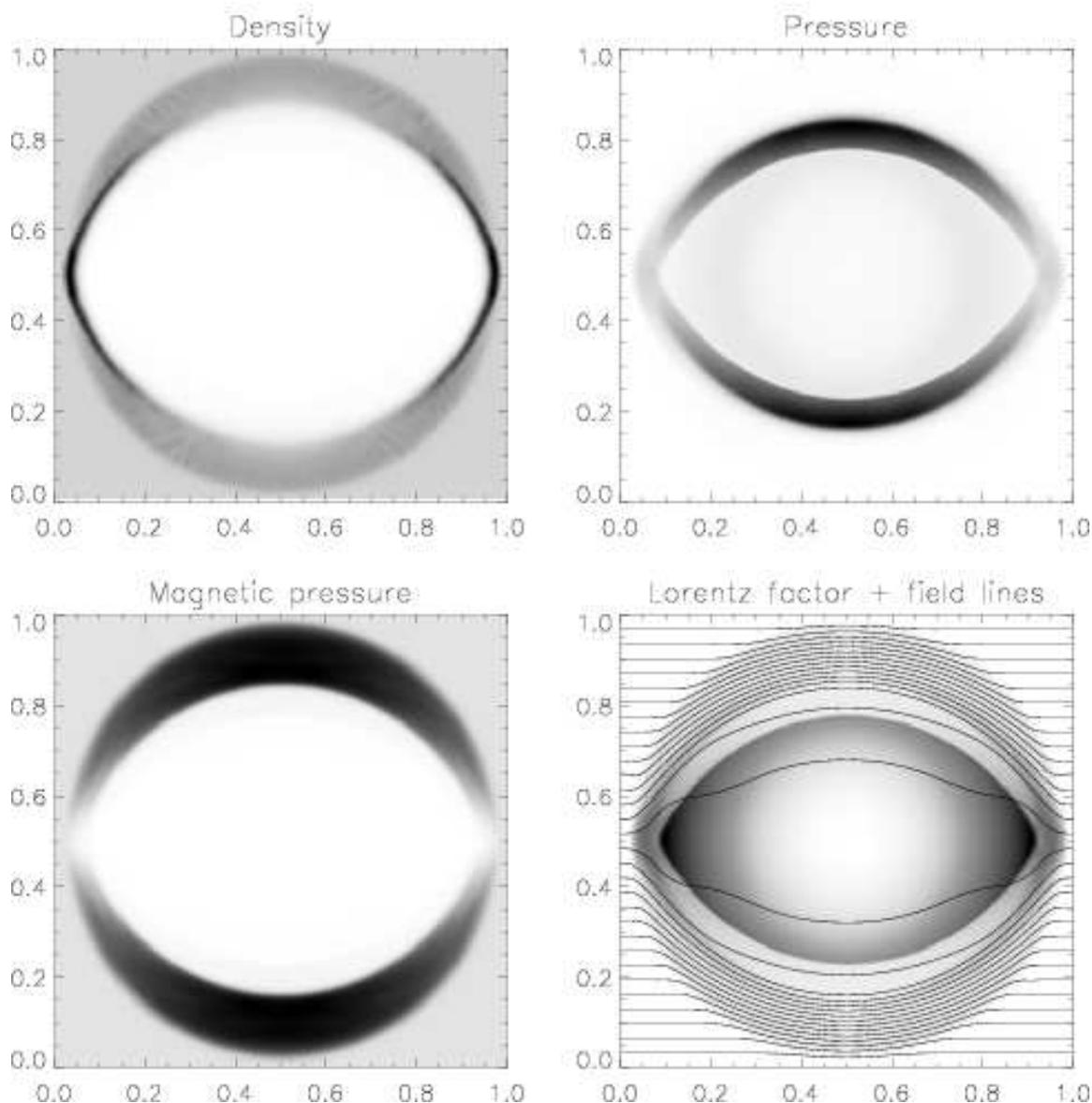}}}
\caption{A RMHD 2-D strong explosion with a pressure jump as high as $10^5$.
The resolution is $N_x=N_y=250$ grid points, and the base multidimensional
scheme employing HLL solver and MM limiter is used.
Grayscale levels are displayed for density, kinetic pressure,
magnetic pressure, and Lorentz factor (together with field lines), 
where $5.36\times 10^{-3}\leq\rho\leq 5.79$,
$0.0< p\leq 45.2$, 
$4.32\times 10^{-2}\leq p_m\leq 72.2$,
and $1.0\leq\gamma\leq 4.35$.
}
\end{figure*}

\begin{figure*}
\centerline{\resizebox{.85\hsize}{!}{\includegraphics{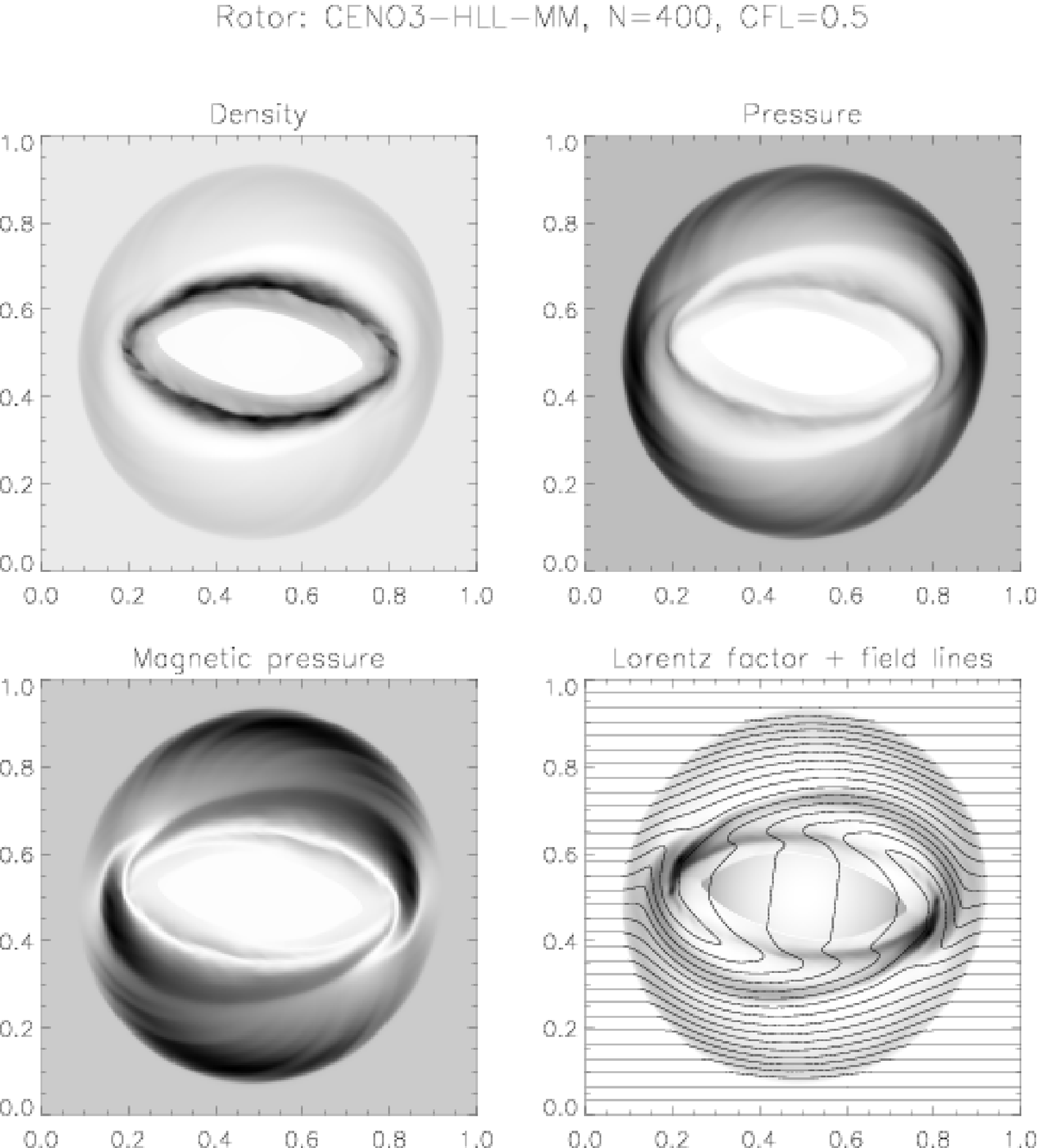}}}
\caption{The relativistic analog of the {\em rotor} test, with
an initial maximum Lorentz factor of about 10.
A high resolution ($N_x=N_y=400$ grid points) simulation is
shown, with the same numerical settings as in Fig.~4.
Grayscale levels are displayed for density, kinetic pressure,
magnetic pressure, and Lorentz factor (together with field lines), 
where $0.35\leq\rho\leq 8.19$,
$5.31\times 10^{-3}\leq p\leq 3.88$, 
$3.77\times 10^{-4}\leq p_m\leq 2.43$,
and $1.0\leq\gamma\leq 1.79$.
}
\end{figure*}

The numerical verification of the code is reported here in three separate 
sub-sections: 1-D shock tubes are presented in the first, some 
2-D test simulations of astrophysical interest are shown in the second,
while the third subsection is devoted to more quantitative tests concerning
code convergence on smooth fields in both 1-D and 2-D.

Since in 1-D RMHD the solenoidal constraint is automatically satisfied 
and transverse magnetic field components behave essentially
like the other conservative variables, the shock tube tests shown
here just illustrate the ability of the code to handle degenerate
cases (where the system is no longer strictly hyperbolic due to the
coincidence of two or more eigenvalues) and to separate the various
Riemann discontinuities or rarefaction waves, which are more numerous in 
the magnetized case (up to seven) rather than in the fluid case
(just three).
On the other hand, multidimensional tests truly prove the robustness
of the code and its accuracy in preserving $\nabla\cdot\vec{B}=0$ in
time, thus avoiding the onset of spurious forces due to the
presence of numerical magnetic monopoles. The solenoidal
constraint is preserved within machine accuracy, like for all
CT schemes, and therefore the spatial distribution of $\nabla\cdot\vec{B}$
and its evolution in time will not be shown (however, see LD for proofs
in classical MHD tests).

For all the simulations that we will show, the scheme described
above is applied without any additional numerical viscosity term, 
which are instead introduced by both KO and BA in order
to stabilize their Roe-type codes. The numerical parameters that may 
be changed in our simulations are just the CFL number 
(here always $c=0.5$),
the reconstruction slope limiter (the Monotonized Centered, MC, or the 
most smearing MinMod, MM, for the multidimensional highly relativistic
tests shown below; see Paper~I for the precise definition of these limiters),
the order of the reconstruction (third, with CENO routines, whenever
possible), and the central-type averaged Riemann solver (we always
use the HLL solver). Concerning this last point, we have verified that 
HLL and LLF actually give almost identical results in all simulations.
This is easily explained: in RMHD either the sound speed or the Alfv\'en
speed are often high, especially at shocks where upwinding becomes
important, so that $\alpha^+\sim\alpha^-\sim\alpha\sim 1$
and the two fluxes tend to coincide.

\subsection{One-dimensional tests}

Shock-tube Riemann problems are not really suited for high order
shock capturing codes, because oscillations may easily appear
near discontinuities.
This is especially true when the reconstruction is not applied
to characteristic waves, because the various contributions
cannot be singled out and, for example, it is impossible to steepen
numerically contact or Alfv\'enic discontinuities.
However, we will see here that the base third order CENO3-HLL-MC scheme
is able to treat this kind of problems reasonably well, usually achieving
similar or even better accuracy than characteristics-based second order
schemes.
In Table~1 the parameters for the initial left (L) and right (R) states 
of the proposed Riemann problems are reported (in all cases $v_y=v_z=0$,
$\Gamma=5/3$ and $t=0.4$). These are the same tests as in BA, except
the first where $\Gamma=2$ was used. Moreover, for ease of comparison,
the same resolution used in BA, $N=1600$ grid points, is employed.

\begin{table}
\begin{center}
\begin{tabular}{lrrrrrr}
\hline
Test & $\rho$ & $v_x$ & $p$ & $B_x$ & $B_y$ & $B_z$ \\
\hline\hline
1~L & 1.0 & 0.0 & 1.0 & 0.5 & 1.0 & 0.0 \\
1~R & 0.125 & 0.0 & 0.1 & 0.5 & -1.0 & 0.0 \\
\hline
2~L & 1.0 & 0.0 & 30.0 & 5.0 & 6.0 & 6.0 \\
2~R & 1.0 & 0.0 & 1.0 & 5.0 & 0.7 & 0.7 \\
\hline
3~L & 1.0 & 0.0 & 1000.0 & 10.0 & 7.0 & 7.0 \\
3~R & 1.0 & 0.0 & 0.1 & 10.0 & 0.7 & 0.7 \\
\hline
4~L & 1.0 & 0.999 & 0.1 & 10.0 & 7.0 & 7.0 \\
4~R & 1.0 & -0.999 & 0.1 & 10.0 & -7.0 & -7.0 \\
\hline
\end{tabular}
\end{center}
\caption{Constant left (L) and right (R) states for the Riemann
problems.}
\label{Table1}
\end{table}

The results relative to the first test are shown in Fig.~1.
This is the relativistic extension of the classic Brio \& Wu (1988)
test, where a {\em compound}, or intermediate, shock wave is formed.
There is still a debate going on about the reality of such structures,
invariably found by any shock-capturing code but not predicted by
analytic calculations (e.g. Barmin et al. 1996; Myong \& Roe 1998).
However, it is not our intention to contribute to that debate, here
we just want to show that our third order reconstruction with
Monotonized Center slope limiter gives better accuracy for both
the left-going intermediate shock and the right-going slow shock,
in comparison to the second order scheme of BA (which employs
a MinMod limiter on magneto-sonic shocks and a special steepening 
algorithm for linearly degenerate characteristic variables, i.e. Alfv\'enic
and contact discontinuities, actually switched off for compound waves). 
As we can see, the total absence of characteristic waves decomposition in 
our code does not prevent at all the sharp definition of discontinuities.
Moreover, oscillations due to high order reconstruction and to the
use of the most compressive MC limiter, evident in the $v_x$ profile, 
are kept at a very low level, while, at the same time, transitions
between constant states and rarefaction waves are rather sharp.

A couple of blast wave examples are shown in Fig.~2, again taken
from BA, the first with a moderate pressure jump and the second
with jump as large as $10^4$, producing a relativistic flow with
a maximum Lorentz factor of $\gamma\simeq 3.4$. Also in these cases
our results are basically equivalent to those in BA, in spite
of some small spurious overshoots (more apparent in the $\rho$
profile in the upper panel and in the $\gamma$ profile of the second panel),
due to the compressive limiter, and of a rather poorly resolved contact 
discontinuity (in the first test, in the second the density peak
is far too narrow to recognize it), due to the fact that we cannot 
steepen it artificially because our component-wise reconstruction.
Again, oscillations are nearly absent and rarefaction waves are
very well defined. The performances on this kind of tests mainly
depend on the limiter choice and on the reconstruction order,
so both accuracy and numerical problems are similar to those
already shown in the RHD case.

Finally, in Fig.~3 we show the fourth test proposed by BA, which
is the magnetic extension of the shock reflection problem of Paper~I.
To reduce post-shock oscillations, more evident in the pressure profile,
we have run this test at second order, thus as in BA; 
in spite of this our slow shocks are better
resolved and the wall heating problem produces a lower dip in the
density profile at $x=0.5$. We have also run this test by using
highly relativistic flows with $\gamma\simeq 224$, as in Paper~I, and
we have not met any particular problem. The good performance of
our code in this last test, in its second order version, is due to
the use of the HLL solver which is not based on the definition
of an intermediate state, based on the left and right reconstructed
quantities, for the definition of characteristic speeds, as it is
done in usual Roe-type solvers.

\subsection{Multidimensional tests}

For truly multidimensional RMHD tests, analytic solutions are not
available and so the verification of the code must be done at a rather
qualitative level. Here we will present a cylindrical blast explosion,
a cylindrical rotating disk, both in 2-D Cartesian coordinates with
a uniform magnetized medium, and the same astrophysical jet of Paper~I
in cylindrical coordinates, now propagating in a magnetized background.
The only other 2-D RMHD code for wich extensive numerical verification
is available in the literature is KO,
where two blast explosions and a Cartesian 2-D jet were tested, in
addition to some simulations of simple 1-D waves and shocks on a 2-D grid
which will not be repeated here.

In the first test we use the standard $[0,1]\times [0,1]$ Cartesian grid,
here with a resolution of $N_x=N_y=250$ grid points,
and we define an initially static background with $\rho=1.0$, $p=0.01$
and $B_x=4.0$. The relativistic flow comes out by setting a much higher
pressure, $p=10^3$, within a circle of radius $r=0.08$ placed at the
center of the domain. Here we use $\Gamma=4/3$ to reduce plasma
evacuation at the center.
In Fig.~4 we show the situation at $t=0.4$, when the flow has almost
reached the outer boundaries. The flow speed reaches its maximum value
along the $x$ direction, $\gamma_\mathrm{max}\simeq 4.35$, because
the expansion of the blast wave is not slowed down by the presence
of a transverse magnetic field, as it happens along $y$ where field lines
are squeezed producing the highest magnetic pressure.
Magnetized cylindrical blast wave are a nice tool to investigate the
behavior of the plasma, and the robustness of the code, in a variety
of degenerate cases (see KO for a detailed description of the various
types of shocks involved). In this simulation we can see that, despite
the absence of appropriate Riemann solvers handling the degeneracies,
our code gives smooth and reasonable results in all directions.
If we compare with the results shown by KO, we may see that in spite
of the absence of additional artificial resistivity and of the 
smoothing of the initial structure (both included in KO), our results
are rather smooth. Only low-level noise in the density may be seen
in the expanding density shell near the diagonals, reminiscent of
the {\em numerical artifacts} found by KO in a run without resistivity.
These errors are possibly due to the use of Cartesian geometry,
since numerical errors on the independent $v_x$ and $v_y$ reconstructions
are the largest precisely along diagonals. 

Another point raised by KO is the possibility of non-strict total
energy conservation even in CT upwind MHD schemes, since magnetic
field components are stored and evolved at different locations
rather than at cell centers where fluid variables are defined.
However, if the total energy, which obviously is a conservative variable, 
is not re-defined in order to prevent unphysical states, it must be globally 
conserved algebraically.
We have checked that in this 2-D test the total energy is conserved
within machine accuracy, as expected. 
In our opinion, the results found by KO in his set of analogue tests, where
the total energy is shown to decrease in time (up to a value as large as 3\% 
in the intermediate case, see his Fig.~12), are mainly due to the presence
of a non-consistent treatment of the artificial resistivity, which is absent
in our code.  

The same numerical parameters, but with a higher resolution
($N_x=N_y=400$), are employed in the second simulation, here adapted to the 
relativistic case from the classical MHD one (Balsara \& Spicer 1999b;
LD; T\'oth 2000). A disk of radius $0.1$ with higher density, $\rho=10$, 
rotating at high relativistic speed, $\omega=9.95\Rightarrow
\gamma_\mathrm{max}\simeq 10.0$, the {\em rotor}, is embedded
in a static background with $\rho=1.0$, $p=1.0$ and $B_x=1.0$ ($\Gamma=5/3$).
In Fig.~5 the complicated pattern of shocks and torsional Alfv\'en waves
launched by the rotor may be seen at the usual output time $t=0.4$,
when the central field lines are rotated of an angle of almost $90^\circ$.
This magnetic braking slows down the rotor, whose maximum Lorentz
at the output time is just $\gamma=1.79$. Note how the initial high
density central region has been completely swept away: the density
has now its minimum ($\rho=0.35$) at the center and the material
has gone to form a thin oblong-shaped shell.
In spite of the presence of very strong shear flows (again, no smoothing
is applied to the disk boundary in the initial conditions), it appears
that our central high order HLL solver is good enough both in providing
high accuracy on smooth waves and in preventing numerical oscillations
at shocks. The turbulent aspect of the high density shell should be
due to the nonlinear evolution of shear-flow instabilities, since
its position coincide with the transition layer where the flow changes
its direction from tangential to radial.

\begin{figure}
\resizebox{\hsize}{!}{\includegraphics{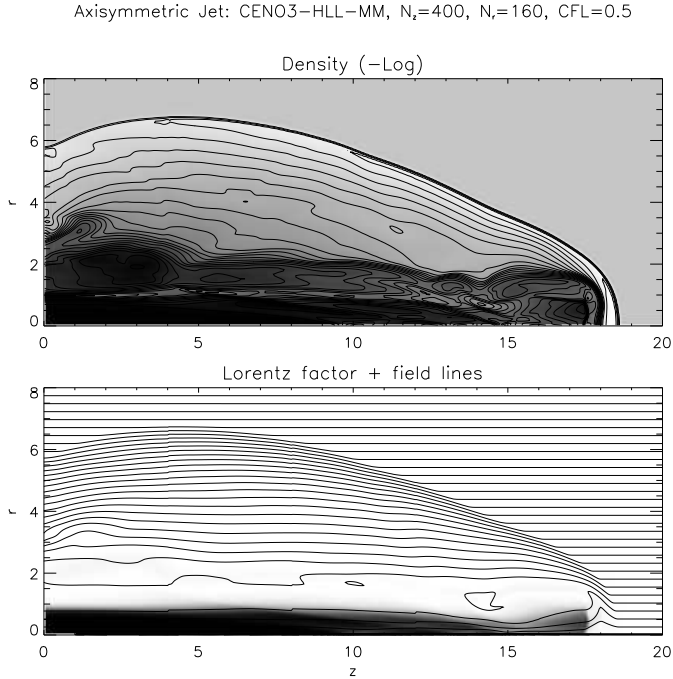}}
\caption{Magnetized axisymmetric jet simulation in cylindrical
coordinates $z$ and $r$, with $N_z=400$ and $N_r=160$ 
(20 grid points per inlet radius are employed). The evolution
is shown here for $t=35$, with the same settings as in Fig.~4 and 5.
In the top panel grey-scale shades
and contours are displayed for $-Log_{10}(\rho)$, with 
$\rho_\mathrm{min}\simeq 0.095$ (black) and 
$\rho_\mathrm{max}\simeq 38.6$ (white).
In the bottom panel Lorentz factor and the magnetic field lines
are displayed together, where $\gamma_\mathrm{max}\simeq 7.14$ (black).}
\end{figure}

Finally, for a truly astrophysical application and as a test of the 
ENO-CT scheme in a non-Cartesian geometry, we simulate the
propagation of a relativistic axisymmetric jet in cylindrical
coordinates. The initial settings are taken the same as in Paper~I,
for ease of comparison with the non-magnetized case.
These are a domain $[0,20]$ along $z$ ($N_z=400$) and
$[0,8]$ along $r$ ($N_r=160$), corresponding to a common
resolution of 20 grid points per inlet radius, a static background plasma 
with $\rho=10.0$, $p=0.01$, $B_z=0.1$ ($\Gamma=5/3$), and jet parameters
of $v_z=0.99$ and $\rho=0.1$, while pressure and magnetic fields are
the same as in the external medium, corresponding to a density ratio
$\eta=1/100$ and to a relativistic Mach number $M=\gamma v/\gamma_{c_s}c_s
=18.3$ and to a relativistic Alfv\'enic Mach number 
$M_A=\gamma v/\gamma_{c_A}c_A=24.3$, where $c_s^2=\Gamma p/w_\mathrm{tot}$
and $c_A^2=B^2/w_\mathrm{tot}$.
Boundary conditions are reflective at the axis and extrapolation
is assumed across the other boundaries. In the region $0<z<1$, $0<r<1$,
the initially smoothed jet values are kept constant in time.

The evolution is shown in Fig.~6 at $t=35$, where
the density logarithm, the Lorentz factor and the magnetic field lines
are displayed. Note that the head of the jet moves faster than 
in the non-magnetized case, because of the confinement due to 
the compressed magnetic field lines (initial equipartition is assumed,
so $B^2=p$) that also reduces the extension of the {\em cocoon}
and stabilizes Kelvin-Helmoltz instabilities, so nicely defined
in Paper~I. Additional reasons for this latter aspect are of numerical 
type: the use of MM rather than MC slope limiter and the higher
numerical viscosity, due to the magnetic contribution in the fast 
magneto-sonic speeds, introduces extra smoothing of contact discontinuities.

As we can see from this set of 2-D examples (the 3-D case does not present 
any additional difficulty), our code is able to obtain similar results 
to those shown in KO. Like in the fluid case, we have found that the
higher order of the scheme can compensate the lack of characteristic
waves decomposition. Even the physical limits that the code is capable
to cope with look very similar to both KO and Koide et al. (1996), 
essentially because of errors in reconstruction of multi-dimensional
vectors. In fact, separate 1-D reconstruction
on vector components may easily produce unphysical states, for example
$v^2$ may be greater than one or the errors on $B^2$, needed for fluxes
and Alfv\'enic or magneto-sonic speed calculations, may be too large leading 
to states with superluminal characteristic modes. While the former problem
may be cured by eliminating the reconstruction in some cases, the latter
is more difficult to prevent. Another problem, common to classical MHD
(see Balsara \& Spicer 1999a),
appears in situations of low-beta plasma, where $\beta=p/B^2$.
When the magnetic field is too strong, the pressure is derived numerically
as a difference of two very large numbers (the total energy and the
magnetic pressure), so it may even become negative.
In our code, when the routine described in Sect.~2.3 still manages to
find a solution for $v^2$ but then negative pressures are found,
we reset $p$ to a small value ($10^{-6}$). The lowest value of the
plasma beta that the code is able to handle, when relativistic 
flows are present, appears to be around $10^{-3}-10^{-4}$. Typical
critical situations are strong rarefactions, as in the 2-D blast
wave presented above.

\begin{figure}
\resizebox{\hsize}{!}{\includegraphics{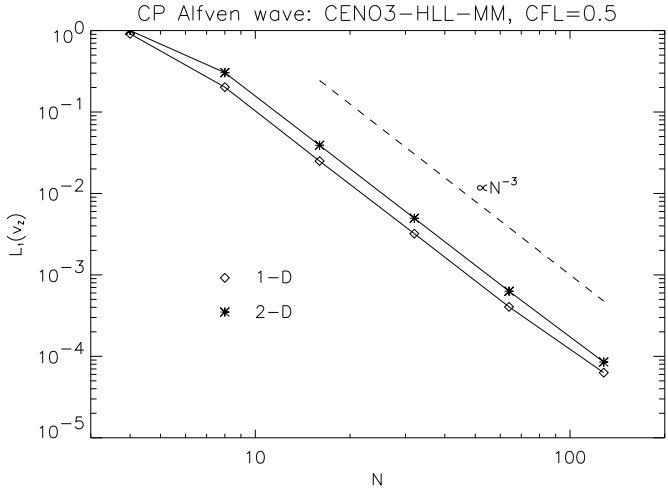}}
\caption{Convergence test for the 1-D and 2-D CP Alfv\'en wave problem.
Relative $L_1$ errors on $v_z$ are shown in logarithmic scale. Note that
the code soon achieves third order convergence in both cases. The errors
in the 2-D case are larger because two periods fit on the main diagonal.
The last point in the 1-D case refers to a slightly larger error than
expected for perfect third order accuracy, because relativistic effects
begin to appear on the wave profile and we are no longer comparing with 
the correct solution.}
\end{figure}

\subsection{Convergence tests}

The tests shown in the previous subsections were mainly devoted to show
the robustness of the code on highly relativistic flows and shocks, thus
also proving the useful property of the CENO3 reconstruction algorithm
to reduce itself to lower orders near discontinuities, avoiding oscillations 
typical of high-order central-type schemes.
In the present subsection we check the high-order accuracy properties of the 
CENO3 interpolation routines on smooth fields: in cases where discontinuous 
features are absent, these algorithms should actually achieve third order
accuracy. 
However, the reader might wonder whether the overall RMHD scheme, which
is based on a rather complicated sequence of CENO3 reconstruction and
derivation routines, especially in the multidimensional case, is able
to preserve global third order accuracy properties in both time and space.

To this porpouse let us study the propagation of relativistic circularly
polarized (CP) Alfv\'en waves. In the limit of small amplitudes the
total magnetic field strength is preserved in time, the Alfv\'en speed
is given by $B_0/\sqrt{w_\mathrm{tot}}$ and the relation between velocity
and magnetic fluctuations reduce to $\delta\vec{v}=\pm\delta\vec{B}
\sqrt{w_\mathrm{tot}}$, similarly to the classical MHD case.
Define now the various quantities in a generic cartesian reference frame
($\xi,\eta,\zeta$) as $\rho=1$, $p=0.1$, $B_\xi=B_0=1$, $v_\xi=0$, and
\be
B_\eta=-A\cos(2\pi\xi),~~~ B_\zeta=-A\sin(2\pi\xi),
\ee
where we have taken $A=0.01$. In the 1-D case we simply have $(\xi,\eta,\zeta)
=(x,y,z)$, whereas in the 2-D case we consider propagation along the $x=y$
direction, so that $(\xi,\eta,\zeta)=((x+y)/\sqrt{2},(-x+y)/\sqrt{2},z)$.
In both cases $[0,1]$ intervals and periodic boundary conditions have been 
assumed ($B_0=\sqrt{2}$ in the 2-D case in order to satisfy these conditions).

The convergence can be proved by measuring relative errors of a certain
quantity, $v_z$ in our case, at different resolutions, where the error has been 
evaluated as the $L_1$ norm of the numerical solution after one period $T$, 
compared to the initial settings:
\be
L_1(v_z)=\frac{\sum_{ij}|v_z(x_i,y_j,t=T)-v_z(x_i,y_j,t=0)|}
{\sum_{ij}|v_z(x_i,y_j,t=0)|}.
\ee
In Fig.~7 the errors are plotted in both 1-D and 2-D cases as a function
of the number of grid points employed $N=N_x=N_y$, in logarithmic scale.
As expected, third order accuracy is achieved already in low resolution
runs. The base scheme employed is CENO3-HLL-MM, which gives the smoothest
profiles, more appropriate to wave-like features. However, we have also
tested the sharper MC limiter: third order accuracy is globally preserved, 
but behaviour of the relative errors as function of the resolution appears
to be more oscillatory, probably due to artificial compression
that tends to sharpen somehow even sinusoidal waves (see Fig.~5 of Paper~I).

\section{Conclusions}

The shock-capturing 3-D MHD scheme of Londrillo \& Del Zanna (2000) is
applied to the special relativistic case, thus extending the code 
for relativistic gasdynamics described in Del Zanna \& Bucciantini (2002),
Paper~I, to the magnetic case. This is the first higher than second order 
(third) upwind scheme developed for RMHD, to which high resolution 
Godunov-type methods have started to be applied only very recently.
Instead of defining complicated linearized Riemann solvers, usually
based on reconstructed characteristic fields, our scheme just uses
the local fastest characteristic velocities to define a two-speed 
HLL-type Riemann solver.
Moreover, reconstruction is applied component-wise, thus time-consuming
spectral decomposition is avoided completely, in the spirit of
the so-called central schemes. This is of particular importance in both
MHD and RMHD, since we do not need to worry about ubiquitous degenerate cases, 
usually handled in Roe-type schemes by adding artificial numerical viscosity.

A main feature of our code is the correct treatment of the solenoidal
constraint, which is enforced to round-off machine errors by extending
the constraint transport (CT) method, originally developed for the induction
equation alone, to the overall RMHD system: the flux functions are correctly 
defined by using the staggered magnetic field components, thus avoiding the onset 
of monopoles even at discontinuities. 
It is important to notice that, in order to obtain such results,
$\nabla\cdot\vec{B}$ must be kept equal to zero at cell centers and it
must be calculated by using the same staggered components which are
evolved in time and the same discretizations applied to flux derivatives 
(see T\'oth, 2000, for examples where these properties do not apply).
Moreover, numerical fluxes based on four-state reconstructed quantities
are defined for the induction equation and here applied to a two-speed 
central-upwind solver for the first time.
In our opinion, to date our method is the only consistent application of CT 
to an upwind scheme for mixed systems of hyperbolic and Hamilton-Jacobi equations, 
like MHD and RMHD. 

Particular attention has been also devoted to the numerical method
needed to derive primitive variables from the set of conservative ones.
The $5\times 5$ system of nonlinear equation is reduced to just a 
{\em single} equation for the square of the velocity. 
This is then solved via a Newton-Raphson iterative root-finding algorithm, 
and analytical expressions are provided for the function whose 
zeroes are looked for and for its first derivative.
This procedure is extremely efficient and robust, and may be
used in all RMHD codes, regardless of the numerical scheme employed.

The code is verificated against 1-D shock tube tests and 2-D problems,
even in non-Cartesian geometries,
showing accurate results and non-oscillatory profiles. The code is very
robust within the limits imposed by the intrinsic numerical precision,
which for multidimensional relativistic flows appear to be $\gamma\sim
10-20$ ($\gamma>200$ is reached in 1-D calculations) and $\beta=p/B^2\sim
10^{-4}-10^{-3}$. These limits seem to be common to all other existing
RMHD codes, and are essentially due to the fact that physical states
become undistinguishable in the ultra-relativistic regime (e.g. all
characteristic speeds collapse onto the speed of light), where even very
small errors on the reconstruction produce fluxes that lead to unphysical
states. Typical situations where the code may fail are strong rarefactions
in a strongly magnetized medium.

Finally, generalized orthogonal curvilinear coordinates are defined in
the code, and presented in the appendix, so our scheme may be easily 
extended to include General Relativity effects with a given metric.

\begin{acknowledgements}
The authors thank the referee, C. Gammie, for his useful comments
and especially for asking us to include the convergence tests.
This work has been partly supported by the Italian Ministry for University 
and Research (MIUR) under grants Cofin2000--02--27 and Cofin2001--02--10.
\end{acknowledgements}

\appendix
\section{Orthogonal curvilinear coordinates}
The equations for special relativistic MHD in a generalized orthogonal 
curvilinear coordinate system $(x^1,x^2,x^3)$ are obtained by assuming 
a (covariant) metric tensor of the form
\be
g_{\alpha\beta}=\mathrm{diag}\{-1,h_1^2,h_2^2,h_3^3\}.
\ee
The first step is to re-define vector and tensor covariant or contravariant 
components as {\em ordinary} components, then spatial differential
operators must be converted in this new coordinate system. The set of 
conservative equations in the divergenge form, Eq.~(\ref{cons_law}), becomes
\be
\frac{\partial \vec{u}}{\partial t} + \frac{1}{h_1h_2h_3}
\sum_{i=1}^{3} \frac{\partial}{\partial x^i}
\left(\frac{h_1h_2h_3}{h_i}\vec{f}^i\right) + \vec{g} = 0,
\ee
where $\vec{u}$ and $\vec{f}^i$ are still those defined in Eq.~(\ref{cons})
and (\ref{flux}), respectively. The source term $\vec{g}$ contains
the derivatives of the metric elements
\be
\vec{g}=\left(\begin{array}{c}
0 \\
h_{12}T_{12}+h_{13}T_{13}-h_{21}T_{22}-h_{31}T_{33} \\
h_{23}T_{23}+h_{21}T_{21}-h_{32}T_{33}-h_{12}T_{11} \\
h_{31}T_{31}+h_{32}T_{32}-h_{13}T_{11}-h_{23}T_{22} \\
0
\end{array}\right),
\ee
where $T_{ij}=w_\mathrm{tot}u_iu_j-b_ib_j+p_\mathrm{tot}\delta_{ij}$
is the stress tensor and where we have defined
\be
h_{ij}=\frac{1}{h_ih_j}\frac{\partial h_j}{\partial x^i}.
\ee

Concerning the magnetic evolution equations, since in our CT scheme
we have assumed $\vec{A}$ as a primary variable, evolved in time by
Eq.~(\ref{At}), the only changes occur in the derivation
of the magnetic field:
\be
B_i=\frac{h_i}{h_1h_2h_3}\sum_{jk}\epsilon_{ijk}
\frac{\partial}{\partial x^j}(h_kA_k),
\ee
which just expresses $\vec{B}=\nabla\times\vec{A}$ in generalized 
orthogonal coordinates.

In the code, the various combinations of the metric elements are
preliminarly calculated and stored on the required grids.
Thus $(h_1h_2h_3)^{-1}$ and the six $h_{ij}$ terms are defined
at grid points $P_{i,j,k}$, $h_2h_3$ and its reciprocal are defined 
at the staggered grid $P_{i+1/2,j,k}$ (similarly for $h_3h_1$ and 
$h_1h_2$), where the corresponding flux and longitudinal divergence-free 
magnetic field component need to be calculated, and finally the
$h_i$ elements are stored on the same grids where $A_i$ and $E_i$
are defined, that is $h_1$ on $P_{i,j+1/2,k+1/2}$ and so on.
Thus, to obtain the derivatives in the above expressions, we just 
need to multiply numerical fluxes and potential vector components
with the corresponding geometrical terms, and then we may proceed
in the usual way.

\end{document}